\documentclass[prc,twocolumn,showpacs,floatfix,superscriptaddress,amsmath,amssymb,nofootinbib]{revtex4-1}

\usepackage{graphicx,color}
\usepackage{wrapfig}
\usepackage{amsmath,amssymb}

\newcommand{\bra}[1]{\langle {#1} |}
\newcommand{\ket}[1]{| {#1} \rangle}
\newcommand{\inproduct}[2]{\langle #1 | #2 \rangle}

\begin{document}


\title{Systematic investigation of low-lying dipole modes\\
using the canonical-basis time-dependent Hartree-Fock-Bogoliubov theory}


\author{Shuichiro~Ebata}
\affiliation{Center for Nuclear Study, University of Tokyo, Wako-shi, 351-0198, Japan}
\affiliation{RIKEN Nishina Center, Wako-shi, 351-0198, Japan}
\author{Takashi~Nakatsukasa}
\affiliation{RIKEN Nishina Center, Wako-shi, 351-0198, Japan}
\affiliation{Center for Computational Sciences, University of Tsukuba,
Tsukuba 305-8571, Japan}
\author{Tsunenori~Inakura}
\affiliation{RIKEN Nishina Center, Wako-shi, 351-0198, Japan}


\date{\today}

\begin{abstract}
Systematic investigations of the electric dipole ($E1$) modes
of excitation are performed using 
the canonical-basis time-dependent Hartree-Fock-Bogoliubov (Cb-TDHFB) theory. 
The Cb-TDHFB is able to describe dynamical pairing correlations
in excited states of nuclear systems. 
We apply the method to the real-time calculation of linear response in
even-even nuclei with Skyrme functionals. 
Effects of shell structure, neutron skin, deformation,
and neutron chemical potential (separation energy) are studied
in a systematic way.
This reveals a number of characteristic features of the low-energy $E1$ modes.
We also find a universal behavior in the low-energy $E1$ modes for heavy neutron-rich isotopes,
which suggests the emergence of decoupled $E1$ peaks beyond $N=82$.
\end{abstract}


\maketitle


\section{Introduction}
\label{sec:introduction}

Elementary modes of excitation provide us with fundamental information
on quantum many-body systems.
A novel structure in the ground state may lead to a new type of
elementary excitations.
In studies of radioactive isotopes,
the low-energy electric dipole ($E1$) modes of excitation,
Pygmy dipole resonance (PDR),
are currently of significant interest.

The giant dipole resonance (GDR) at high frequency is
a fundamental mode of excitation in finite nuclei.
The GDR is universally observed in all nuclei,
from light to heavy, from stable to unstable isotopes,
and exhausts almost 100 \% of the energy-weighted sum rule (EWSR) value.
It is also know to reflect bulk properties of nuclei,
such as the symmetry energy.
In contrast, the PDR is sensitive to structure of an individual nucleus.
Since most of the $E1$ strength is carried by the GDR,
the $E1$ strength of low-energy states is known to be significantly hindered.
The presence of substantial $E1$ strength in the PDR
implies that the structure of the nucleus is significantly different from
our conventional understanding.
Thus, it is important to identify the basic property and the origin of
the PDR.

The low-energy dipole states
have been observed in a number of nuclei; neutron-rich O isotopes 
\cite{Leistenschneider,Tryggestad}, $^{26}$Ne \cite{Gibelin}, stable Ca 
isotopes \cite{Hartmann,Isaak}, $^{56}$Fe and $^{58}$Ni \cite{Bauwens}, 
$^{68}$Ni \cite{Wieland}, $^{88}$Sr \cite{Kaubler}, $^{90}$Zr 
\cite{Schwengner,Iwamoto}, Sn isotopes \cite{Govaert,Adrich,Ozel,Klimkiewicz,Endres10}, $N=82$ isotones 
\cite{Herzberg,Zilges,Volz,Savran,Endres09,Tonchev}, and stable Pb isotopes 
\cite{Chapuran,Ryezayeva,Endres00,Tamii,Poltoratska}.
The observed low-energy $E1$ strength exhausts less than 1 \%\ of the 
Thomas-Reiche-Kuhn (TRK) sum-rule value in stable isotopes,
while it may amount up to about 5 \%\ in neutron-rich nuclei.

Theoretically, the PDR has been commonly investigated using
the (quasi-particle) random phase approximation ((Q)RPA)
\cite{Colo,Ter05,Eba10,Yuksel,VPRL01,PRNV03,Cao,Piekarewicz06,
Liang,PR08,Nakada,Los10,Piekarewicz11,DR11,RPBMC12,Carbone,Peru}.
There are some calculations beyond the (Q)RPA as well
\cite{Sarchi,Tsoneva,Litvinova,Avdeenkov,Gambacurta,TN12}. 
However, even in the (Q)RPA level,
different models often predict different properties.
The nature of the PDR is still elusive with a lack of general agreement. 

In order to clarify properties of the PDR,
we think it important to study a variety of isotopes systematically,
from light to heavy, from spherical to deformed, and from normal
to superfluid nuclei.
Using a technique of the finite amplitude method
\cite{NIY07,INY09,AN11,Sto11,AN13,LNZM13,HKN13},
a systematic investigation for the PDR has been carried out for nuclei
up to $Z=40$ in Ref.~\cite{INY11}, in which
the pairing correlations were neglected.
In this article, we report an extended analysis to heavier region,
taking into account the pairing correlations.
To perform the study, we utilize
the canonical-basis time-dependent Hartree-Fock-Bogoliubov (Cb-TDHFB)
method \cite{Eba10}.
We carry out the systematic calculation for the $E1$ modes
of excitation, with a parallelized computer code for the real-time evolution
of the Cb-TDHFB equations
in the three-dimensional (3D) coordinate-space representation.
The method is able to take a full account of nuclear deformation
of any kind, and treat the paring correlation in a BCS-like scheme,
both in static and in dynamic natures.
The numerical cost of the Cb-TDHFB method is significantly smaller than
that of the normal QRPA calculations \cite{Eba10},
which enables us to perform a systematic analysis for nuclei
in a wide region of nuclear chart.
However, since the BCS-like treatment of the pairing correlation leads to
the neutron gas problem very near the drip line\cite{DFT84},
we limit our studies to nuclei with the neutron separation energies
larger than $2.0$ MeV.

The nature of the low-lying modes of excitation in nuclei is
still an open issue.
Especially, the $E1$ modes in neutron-rich nuclei attract much attention
with a simple geometrical picture that the excess neutrons in the nuclear
surface region oscillate against the core of the nucleus.
This interpretation of the PDR as the ``neutron-skin mode''
was originally discussed using hydrodynamical models \cite{MDB71,SIS90,INW92}. 
Then, naturally, one may expect that the collectivity of the low-energy $E1$
transitions has a certain connection to a property of the excess neutrons
and the neutron skin thickness.
The correlation between the $E1$ collectivity of PDR and the neutron skin
thickness has been microscopically studied using the (Q)RPA \cite{Piekarewicz06,RN10,INY11}.
Similar analysis has been performed for
the correlation between the $E1$ collectivity of PDR and the isovector
nuclear matter properties, such as the symmetry energy
\cite{Klimkiewicz,RN10,Carbone,INY13}.
The current situation is that, although everybody agrees with the existence of
certain correlation among them, it is not well settled yet how strong
the correlation is.

In this paper, we carry out the systematic linear-response calculations
for $E1$ mode for over 300 kinds of nuclei.
Properties of the PDRs are investigated in terms of
their isotopic dependence as a function of neutron number,
their relation to the neutron skin thickness and the separation energy,
and deformation effect.

The paper is organized as follows. 
In Sec.~\ref{sec:formalism}, we present the Cb-TDHFB method and 
real-time calculation of the $E1$ strength functions. 
The definition of PDR fraction to quantify the low-lying strength is given, and 
the model-space dependence of the numerical results is examined. 
In Sec.~\ref{sec:NND-PDR}, we show variation of the PDR strengths with respect to 
the neutron number and the neutron-skin thickness,
for nuclei in a wide mass region with $Z\leq 50$. 
We also discuss the functional dependence of the PDRs in 
Sn isotopes with the Skyrme parameter sets of SkM$^*$ and SkI3. 
Section \ref{sec:Df-PDR} presents the deformation effects on the PDR. 
The $K=0$ dominance of the PDR fraction is observed for
nuclei with the prolate shapes. 
We examine its relevance to 
the orientation dependence of the neutron-skin thickness. 
In Sec.~\ref{sec:A82-PDR}, we discuss the PDRs of heavy isotopes around $N=82$.
The present study suggests the existence of PDR peaks
are not hindered by the coupling to GDR.
Finally, the conclusion is given in Sec.~\ref{sec:summary}.

\section{Formulation and numerical details} 
\label{sec:formalism}
In this section, we recapitulate the Cb-TDHFB method for
the calculation of strength functions \cite{Eba10}.
Let us first express
the normal and pair densities, ($\rho_{\mu\nu}(t), \kappa_{\mu\nu}(t)$), 
in terms of canonical pairs of states, ($\phi_k(t)$, $\phi_{\bar{k}}(t)$). 
Here, the indices $\mu, \nu$ mean arbitrary single-particle basis. 
\begin{eqnarray}
\label{rho}
&\rho_{\mu\nu}(t)&= \nonumber \\
&&\hspace{-5mm}\sum_k \rho_k(t) \left\{
\inproduct{\mu}{\phi_k(t)}\inproduct{\phi_k(t)}{\nu} 
+\inproduct{\mu}{\phi_{\bar k}(t)}\inproduct{\phi_{\bar k}(t)}{\nu} 
\right\}, \nonumber \\[-2mm] \\
\label{kappa}
&\kappa_{\mu\nu}(t)&= \nonumber \\
&&\hspace{-5mm}\sum_k \kappa_k(t) \left\{
\inproduct{\mu}{\phi_k(t)}\inproduct{\nu}{\phi_{\bar k}(t)}
-\inproduct{\nu}{\phi_k(t)}\inproduct{\mu}{\phi_{\bar k}(t)} \right\}, \nonumber \\[-2mm]
\end{eqnarray}
where $\rho_k(t)$ and $\kappa_k(t)$ are the occupation and
the pair probabilities, respectively,
which can be written as
$\rho_k(t)=|v_k(t)|^2$ and $\kappa_k(t)=u_k(t)v_k(t)$
using the time-dependent $(u,v)$ coefficients
in the canonical representation \cite{RS80}.
All we need to calculate are the time evolution of
$\phi_k(t)$, $\phi_{\bar k}(t)$, $\rho_k(t)$, and $\kappa_k(t)$.
Equations to determine their time evolution will be given in
the following.
Note that the density operators, $(\rho_{\mu\nu}, \kappa_{\mu\nu})$, 
in the left-hand side of Eqs. (\ref{rho}) and (\ref{kappa}) are
matrices with two indices,
while $(\rho_k,\kappa_k)$ in the right-hand side have a single index.

\subsection{Energy density functional}

The energy density functional is given by the sum of
the Skyrme density functional, $E_{\rm Sky}[\rho]$, and
the pairing energy functional, $E_{\rm pair}[\kappa,\kappa^*]$.
\begin{equation}
\label{E_tot}
E_{\rm tot}[\rho(t),\kappa(t),\kappa^*(t)]
= E_{\rm Sky}[\rho(t)] + E_{\rm pair}[\kappa(t),\kappa^*(t)] .
\end{equation}
For the Skyrme energy functional, the SkM$^*$ parameter set is used,
unless otherwise specified.
For the pairing part, we adopt a simple functional of a form  
\begin{eqnarray}
\label{E_G}
E_{\rm pair}[\kappa(t),\kappa^*(t)] &=& -\sum_{\tau=n,p} \sum_{k,l>0} G_{kl}^{\tau} \kappa_k^{\tau \ast}(t) \kappa_{l}^{\tau}(t) \nonumber \\
&=&-\sum_{\tau=n,p} \sum_{k>0} \kappa_k^{\tau \ast}(t) \Delta_{k}^{\tau}(t),
\end{eqnarray}
where the gap parameter $\Delta_k(t)$ are given by
\begin{equation}
\Delta_{k}^{\tau}(t)= \sum_{l>0} G_{kl}^{\tau} \kappa_{l}^{\tau}(t) . 
\end{equation}
Here, $G_{kl}^\tau=g^\tau f(\varepsilon_k^0) f(\varepsilon_l^0)$ and the
constant $g$ is determined by the smoothed pairing method \cite{Bra72}. 
The cut-off function $f(\varepsilon_k^0)$ depends on the
single-particle energy of the canonical state $k$
at the Hartree-Fock plus BCS (HF+BCS) ground state.
The $f(\varepsilon)$ is written as 
\begin{equation}
f(\varepsilon)=\left( 1+\exp\left[ 
\frac{\varepsilon - \epsilon_{\rm c}}{0.5 \mbox{ MeV}} \right]\
 \right)^{-1/2} \theta (e_{\rm c}-\varepsilon),
\end{equation}
with the cut-off energies 
\begin{eqnarray}
\epsilon_{\rm c} = \tilde\lambda+5.0\ {\rm MeV},\quad
e_{\rm c} = \epsilon_{\rm c}+2.3\ {\rm MeV} ,
\end{eqnarray}
where $\tilde\lambda$ is the midpoint of the highest occupied level
and the lowest unoccupied level in the Hartree-Fock (HF) state.
Here, the cut-off parameter $e_{\rm c}$ is necessary to prevent
occupation of spatially unlocalized single-particle states.
For neutrons, 
if $e_{\rm c}$ becomes positive, we replace it by zero \cite{TTO96}.
Because of this limitation of the HF+BCS,
we restrict our study to nuclei with the neutron separation
energy larger than $2.0$ MeV.

\subsection{Cb-TDHFB equations}

Using the energy functional of Eq. (\ref{E_tot}),
we may derive the Cb-TDHFB equations
based on the full TDHFB equation with an assumption that 
the pair potential is ``diagonal'' in the canonical basis.
Although the Block-Messiah theorem\cite{RS80} guarantees the existence of
the canonical form for the TDHFB state at any time, we need
this diagonal assumption of the pair potential in order to keep the
Cb-TDHFB equations in a simple form during the time evolution.
The details are given in Ref.~\cite{Eba10}.
The Cb-TDHFB equations are given by
\begin{subequations}
\label{Cb-TDHFB}
\begin{eqnarray}
\label{dphi_dt}
&&i\frac{\partial}{\partial t} \ket{\phi_k(t)} =
(h(t)-\epsilon_k(t))\ket{\phi_k(t)} , \\
&&i\frac{\partial}{\partial t} \ket{\phi_{\bar k}(t)} =
(h(t)-\epsilon_{\bar k}(t))\ket{\phi_{\bar k}(t)} , \\
\label{drho_dt}
&&
i\frac{d}{dt}\rho_k(t) =
\kappa_k(t) \Delta_k^{\ast}(t)
-\kappa_k^{\ast}(t) \Delta_k(t) , \\
\label{dkappa_dt}
&&
i\frac{d}{dt}\kappa_k(t) =
\left(
\epsilon_k(t)+\epsilon_{\bar k}(t)
\right) \kappa_k(t) +
\Delta_k(t) \left( 2\rho_k(t) -1 \right) , \nonumber \\
\end{eqnarray}
\end{subequations} 
where the parameter $\epsilon_k(t)$ are taken as
$\epsilon_k(t)\equiv\bra{\phi_k(t)}h(t)\ket{\phi_k(t)}$,
to minimize the phase change of the canonical states.
In this article, we adopt the natural unit, $\hbar=c=1$.
The single-particle Hamiltonian $h(t)$ is defined as usual,
a derivative of the
energy functional $E[\rho,\kappa]$ with respect to the density,
$h_{\mu\nu}(t)\equiv {\partial E}/{\partial\rho_{\nu\mu}}$

These are the basic equations to determine the time evolution of
the canonical states, $\ket{\phi_k(t)}$ and $\ket{\phi_{\bar k}(t)}$,
their occupation $\rho_k(t)$, and pair probabilities $\kappa_k(t)$. 
We use the 3D Cartesian coordinate-space representation for
the canonical states,
$\phi_k(\vec{r},\sigma;t)=\inproduct{\vec{r},\sigma}{\phi_k(t)}$
with $\sigma=\pm 1/2$. 
Thus, each canonical state is represented by
three discrete indexes $(i_x,i_y,i_z)$ for the 3D mesh points;
$(x,y,z)=(i_x,i_y,i_z)\times d$.

\subsection{Real-time calculation of strength functions}
\label{sec:RT-S}

The $E1$ strength function in the linear response
can be obtained in the real-time calculation
with a perturbation of the $E1$ field.
The ground state is prepared by the HF+BCS calculation using the
imaginary-time method.
Then, we add a weak impulse external field 
$V_{\rm ext}(\vec{r},t)=-\eta \hat{F}_K(\vec{r}) \delta(t)$ to start the time evolution.
Here, $\hat{F}_K(\vec{r})$ is the $E1$ operator with
recoil charges,
\begin{equation}
\label{E1}
\hat{F}_K(\vec{r})=\begin{cases}
  (Ne/A) rY_{1K}(\hat{r}) & \text{for protons} \\
- (Ze/A) rY_{1K}(\hat{r}) & \text{for neutrons}
\end{cases} .
\end{equation}
Thus, at time $t=0+$, the initial state of the
time evolution is given by
$\phi_k(\vec{r},\sigma;t=0+) = e^{i\eta \hat{F}_K(\vec{r})} \phi_k^0(\vec{r},\sigma)$,
$\rho_k(t=0+)=\rho_k^0$, and $\kappa_k(t=0+)=\kappa_k^0$,
where the superscript ``0'' indicates the quantities at the ground state.
Then, the expectation value of $\hat{F}$ is calculated as a function of
time, 
\begin{eqnarray}
&&{\cal F}_K(t)\equiv \nonumber \\
&&\int d\vec{r} \left\{ (Ne/A) rY_{1K}(\hat{r})  \rho_p (\vec{r},t)
                                -(Ze/A) rY_{1K}(\hat{r})  \rho_n (\vec{r},t) \right\} , \nonumber \\
\end{eqnarray}
where $\rho_\tau(\vec{r},t)=\sum_{k\in\tau} \rho_k(t) \sum_\sigma
|\phi_k(\vec{r},\sigma;t)|^2$.
The parameter $\eta$ controls the strength of the external field. 
To calculate the linear response, it should be small enough to validate the linearity. 
Note that the strength function $S(E;E1)$ is independent from
magnitude of the parameter $\eta$ as far as the linear approximation is valid. 
In the present study, we adopt the value of $\eta=3\times10^{-5}$
$(e\ \mbox{fm})^{-1}$
for the $E1$ operator. 
The time step $dt$ is 0.0005 MeV$^{-1}$ and 
the time evolution is calculated up to $T=10$ MeV$^{-1}$.

The $E1$ strength function
is calculated with the following formula\cite{NY05}.
\begin{eqnarray}
\label{S(E1)}
S(E;E1) &\equiv& \sum_{K=-1,0,1}
\sum_{n}|\bra{n} \hat{F}_K \ket{0}|^{2}\delta (E-E_n) \nonumber \\
&=& -\frac{1}{\pi\eta}{\rm Im}{\cal F}(E), 
\end{eqnarray}
where $E_n$ is the excitation energy of the state $\ket{n}$
and ${\cal F}(E)$ is defined by
\begin{equation}
{\cal F}(E)=\sum_K 
\int_0^\infty w(t) e^{iEt} {\cal F}_K(t) dt.
\label{F(E)}
\end{equation}
Here, we introduce a smoothing function $w(t)$, for which
the ideal choice would be unity, $w(t)=1$.
However,
because of the finite duration of the numerical time evolution,
the integration must stop at $t=T$.
This requires $w(t)$ to be a function almost vanishing at $t=T$,
and leads to the finite energy resolution,
$\Delta E \sim 2\pi T$.
In the following calculations,
we choose $w(t) \equiv e^{-\Gamma t/2}$, 
where $\Gamma$ is a smoothing parameter and is set to 1 MeV.
This corresponds to the discrete strength smeared with
the Lorentzian function with the width of $\Gamma$.
For spherical nuclei, different $K$ components in Eq. (\ref{S(E1)}) are
all identical.
Thus, the actual numerical calculation is performed only for one
of $K$ values
and the final result is obtained by multiplying it by three.
For axially symmetric nuclei, when the symmetry axis is chosen as $z$-axis,
the $K=1$ and $-1$ components are identical.
The sum of them is simply called ``$K=1$ component'' in the following. 

\subsection{Definition of PDR fraction}

\begin{figure}[h]
\includegraphics[height=80mm, angle=-90]{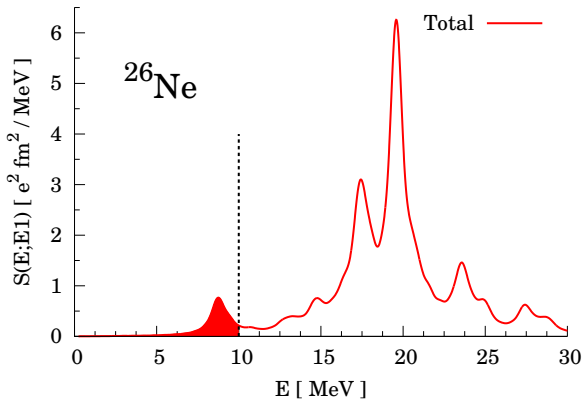}
\caption{Calculated $E1$ strength distribution and PDR strength
(filled area) for $^{26}$Ne. 
A vertical dashed line indicates $E_c=10$ MeV.
}
\label{fig: IVD_Ne}
\end{figure}

To quantify the low-lying $E1$ strength in systematic investigation, 
we use the summed energy-weighted $E1$ strengths in the
low-energy region $m_1(E_c)$.
The ratio of this to the total sum-rule value $m_1$,
\begin{eqnarray}
\label{eq: fracR}
f_\textrm{\sc pdr} = 
 \frac{m_{1}(E_{c})}{m_{1}} \equiv \frac{\int^{E_{\rm c}} E\times S(E1; E) dE}{\int E\times S(E1; E) dE}, 
\end{eqnarray}
is referred to as ``PDR fraction'', hereafter.
The cut-off energy  $E_{c}$ = 10 MeV is adopted in the present calculation. 
Figure \ref{fig: IVD_Ne} shows the $E1$ strength function of $^{26}$Ne
and the low-lying $E1$ strength in the current definition is highlighted.
The PDR fraction, defined in this way, depends on
the $E_c$ and smoothing parameter $\Gamma$.
The present choice of $E_c$ can be reasonably justified for relatively light
nuclei, however, for heavier deformed isotopes, it is difficult to make
a clear separation between the PDR and GDR.
Nevertheless,
although its absolute magnitude depends on $E_c$,
the relative quantities are sensible and useful 
for discussion on the low-lying $E1$ strengths. 

\subsection{Determination of the model space: Box-size dependence}

The canonical orbits $\phi_k(\vec{r},\sigma)$
is represented in the 3D coordinate space discretized with mesh $d$
in the sphere of radius $R$.
It is trivial that
the numerical calculation is easier for a smaller model space
(smaller $R$ and larger $d$).
However, the calculation with too small model space could produce
a qualitatively wrong answer.
Roughly speaking, the box size $R$ (mesh size $d$) determines
the lowest (highest) momentum in the model space.
We here show some results to determine these model-space parameters.

\begin{figure}[h]
   \begin{center}
    \includegraphics[keepaspectratio,width=50mm, angle=-90]{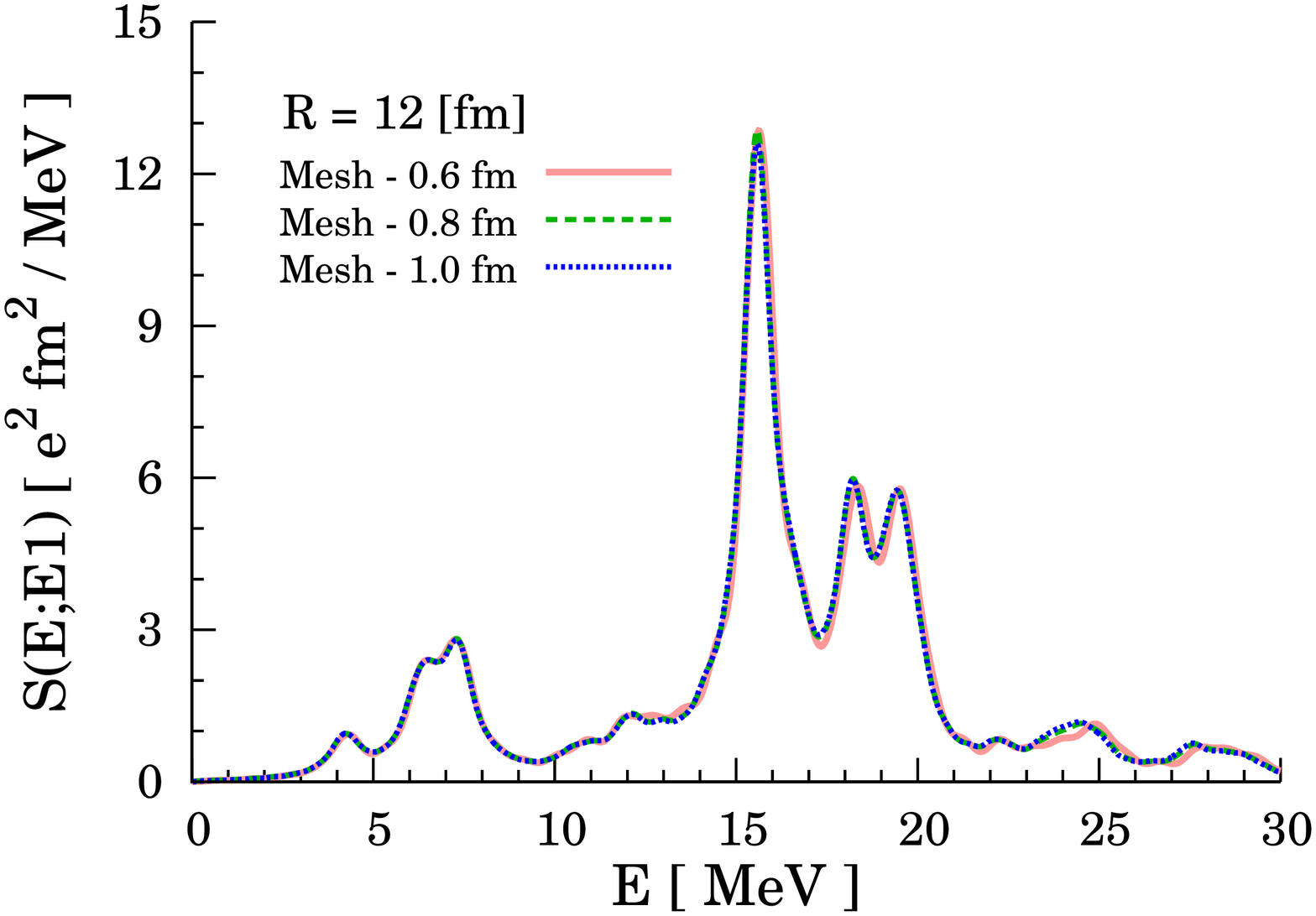}
   \caption{Mesh-size dependence of the strength. These calculations are done in the box of $R=$ 12 fm.}
	\label{fig: MD_S}
    \includegraphics[keepaspectratio,width=50mm, angle=-90]{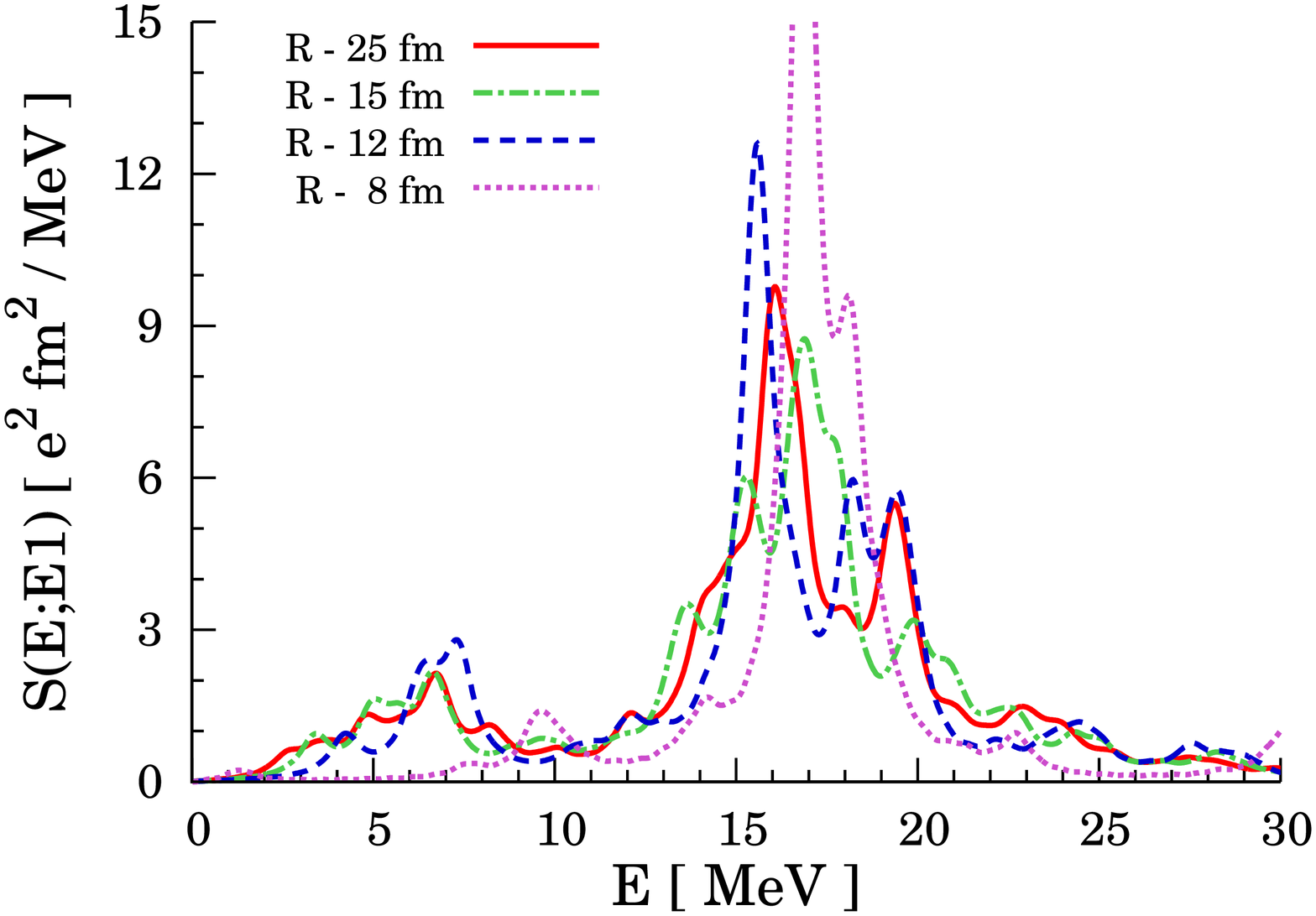}
   \caption{Box-size dependence of the $E1$ strength for $^{50}$S with mesh size of $d=1$ fm.}
	\label{fig: BD_S}
   \end{center}
\end{figure} 
\begin{figure}[h]
\includegraphics[keepaspectratio,width=50mm, angle=-90]{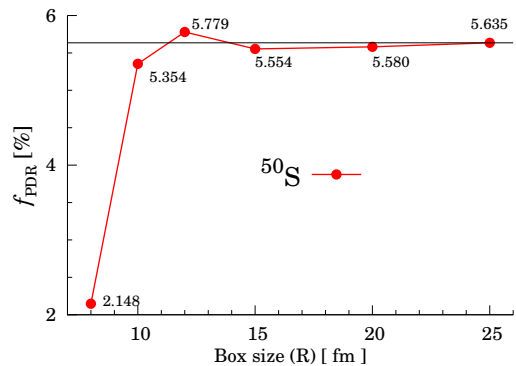}
\caption{Box-size dependence of the fraction of the low-lying $E1$ strength for $^{50}$S.
The thin horizontal line is the result obtained with $R=$ 25 fm.
The values written around symbols are the fraction ratios obtained in each box.
}
\label{fig: fracBD}
\end{figure} 
Figure \ref{fig: MD_S} shows the mesh-size dependence of the $E1$ strength
function for a very neutron-rich nucleus $^{50}$S.
The result is well converged with the use of $d\leq 1$ fm.
In contrast, the box-size dependence is much more significant,
which is illustrated in Fig. \ref{fig: BD_S}.
The solid, dash-dotted, dash and dotted lines correspond to
$R=25$, 20, 12, and 8 fm, respectively.
In the calculation with the smallest space of $R=8$ fm,
the GDR peak is shifted to higher energy and
the low-lying $E1$ peak disappears.
In order to obtain a fully converged result on the
the $E1$ strength functions in $^{50}$S,
we need a very large model space, such as $R\geq 30$ fm.
It should be noted that
this box-size dependence is significantly weaker in stable nuclei,
which was previously investigated in Ref.~\cite{INY09}.

\begin{figure}[h]
  \begin{center}
	 \includegraphics[keepaspectratio,width=50mm, angle=-90]{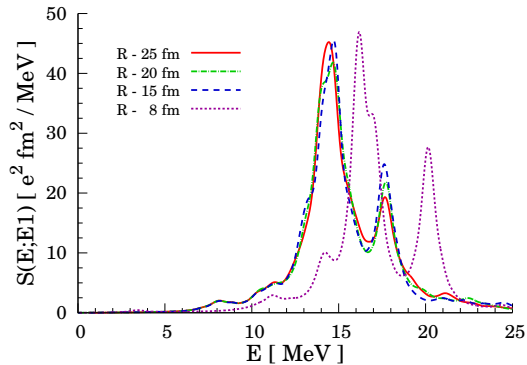}
	\caption{Same as Fig.\ref{fig: BD_S}, but for $^{132}$Sn.}
	\label{fig: BD_SnS}
   \end{center}
\end{figure}

\begin{figure}
   \begin{center}
    \includegraphics[keepaspectratio,width=50mm, angle=-90]{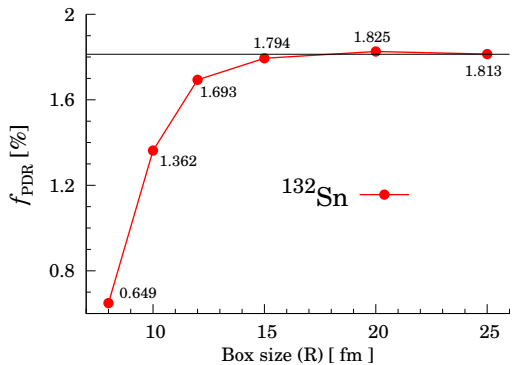}
   \caption{Same as Fig.\ref{fig: fracBD}, but for $^{132}$Sn. 
The thin line is an eye-guide from the result of $R=$ 25 fm case.}
 \label{fig: BD_Snf}
 \end{center}
\end{figure}

It is computationally very demanding to perform a systematic study
in such a large model space of $R\geq 30$ fm.
Thus, we focus our study on the relative and qualitative behaviors of
the low-energy $E1$ strengths.
Figure \ref{fig: fracBD} shows the box-size dependence of the PDR
fraction $f_\textrm{\sc pdr}$ for $^{50}$S. 
The result is approximately converged at $R\geq 12$ fm.
Figures \ref{fig: BD_SnS} and \ref{fig: BD_Snf} show
the box-size dependence of $S(E1;E)$
and $f_\textrm{\sc pdr}$ for a heavier neutron-rich nucleus, $^{132}$Sn. 
The results are almost converged at $R\geq 15$ fm.
From these studies, we here adopt the model space as follows:
$(R,d)=(12,0.8)$ fm for isotopes with $Z\leq 20$,
and $(R,d)=(15,1)$ fm for heavier isotopes with $28\leq Z\leq 50$.
These model spaces lead to
the number of 3D mesh points of  
about 8,000.


We should note that the results for extremely neutron-rich nuclei
in the present paper may not be fully converged, especially for light isotopes.
For instance, we have confirmed that, for Ca isotopes,
the $f_\textrm{\sc pdr}$ value calculated with $R=12$ fm and $R=15$ fm
are identical to each other for $N\leq 36$,
while those for $N>38$ shows some difference.
However, this difference is about 0.6 \% at most,
which does not change the conclusion of the present study.

\section{Evolution of the low-energy $E1$ strengths}
\label{sec:NND-PDR}

In this section, we present how the PDR fraction,
 Eq. (\ref{eq: fracR}), is evolved
with respect to neutron number and proton number.
All results in this section are obtained with the SkM$^*$ parameter set.

\begin{figure}[th]
  \begin{center}
\includegraphics[keepaspectratio,width=50mm, angle=-90]{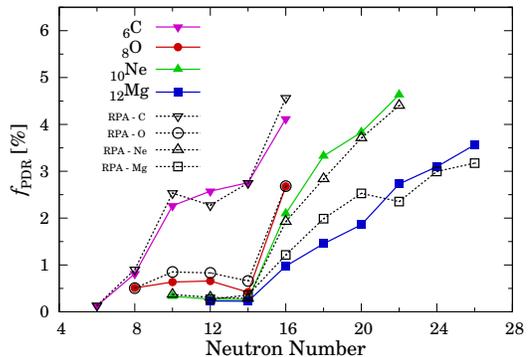}
\caption{
Fraction of the low-lying $E1$ strength $f_\textrm{\sc pdr}$
in Eq. (\ref{eq: fracR}) as functions of the neutron number
for C, O, Ne and Mg isotopes. 
The solid lines with filled symbols show the present results of Cb-TDHFB,
while the dashed lines with open symbols show those of HF+RPA.
}
\label{fig: L1_PDR}
   \end{center}
\end{figure}
\begin{figure}
   \begin{center}
\includegraphics[keepaspectratio,width=50mm, angle=-90]{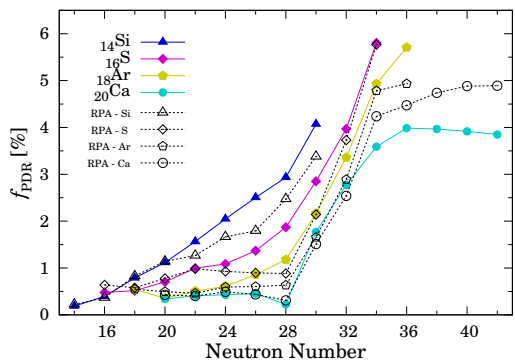}
\caption{Same as Fig.\ref{fig: L1_PDR}, but for Si, S, Ar and Ca isotopes.}
\label{fig: L2_PDR}
  \end{center}
\end{figure}

\subsection{Light nuclei in the neutron-rich side
 (6 $\leq Z\leq$ 20 and $N\geq Z$)}
\label{sec:NND-PDR_L}

First, we show the behavior of the low-lying $E1$ strength
in relatively light even-even isotopes 
with $Z=6 - 20 $ which have the chemical potential
larger than 2 MeV. 
Table \ref{tab: gs_properties} in Appendix 
shows the ground-state properties
calculated with the HF+BCS
for isotopes with $6 \leq Z \leq 20$. 
The results of the HF calculation neglecting the pairing correlation
are also shown for comparison.
There are some difference in the ground-state deformation
between HF+BCS and HF calculations, among which $^{32}$Mg shows
the largest difference, $\beta_2^{\rm HF}=0.35$ and $\beta_2^{\rm HF+BCS}=0.0$.
Nevertheless, for most isotopes, the pairing correlations do not
drastically change the ground-state properties.

Figure \ref{fig: L1_PDR} shows the neutron-number dependence of $f_\textrm{\sc pdr}$ 
for C, O, Ne, and Mg isotopes. 
The solid lines with filled symbols indicate the present results
which can be compared with those of the Hartree-Fock-plus-random-phase 
approximation (HF+RPA) \cite{INY11} 
presented by the dashed lines with open symbols.
The two kinds of calculations produce qualitatively same results.
This confirms that the pairing plays a minor role in the low-energy
$E1$ strength function for these light nuclei.

The isotopes with $Z=8-12$ have $f_\textrm{\sc pdr}$ less than 1.0\%
for $N\leq 14$.
Then, there is a sudden jump in $f_\textrm{\sc pdr}$
at $N=14 \to 16$ on every isotopic chain.
The neutron number $N=16$ corresponds to the occupation of $s_{1/2}$ orbit. 
The important role of the weakly bound $s_{1/2}$ orbit in the low-energy
$E1$ strengths has been discussed in Ref.~\cite{INY11}.
The present result confirms the pairing correlations do not change the
main conclusion.
Note that the neutrons are in the normal phase ($\Delta_n=0$)
for nuclei with $N=16$.
The largest deviation from the HF+RPA result is seen in $^{32}$Mg,
which is due to the large difference in the ground-state deformation. 

The HF+RPA calculation predicts that
the next jump in $f_\textrm{\sc pdr}$ at $N=28 \to 30$ \cite{INY11}.
$N=30$ is corresponding to the occupation of $p_{3/2}$ orbit. 
This is shown in Fig.\ref{fig: L2_PDR},
for S, Ar, and Ca isotopes. 
The qualitative behaviors are identical to those of the HF+RPA calculation. 
For Si, the kink of $f_\textrm{\sc pdr}$ disappears 
because the $N=28$ magicity becomes weak 
in the neutron-rich Si isotopes leading to deformed shapes
in the mean-field calculation (see Tab.\ref{tab: gs_properties}). 
However, for S and Ar isotopes, 
the sudden jump at $N=28 \to 30$ predicted by the HF+RPA is now replaced by
a gradual increase in the slope around $N=28$.
This smooth evolution of the $f_\textrm{\sc pdr}$ is caused by
the fractional occupation probability of 
the special single-particle states, such as $p_{3/2}$ and $p_{1/2}$ orbits,
due to the pairing correlation. 
Again, the occupation of weakly bound orbits with low
orbital angular momenta (low-$\ell$) increases the low-energy $E1$ strength.
Beyond $N=34$, the neutrons start occupying $f_{5/2}$ orbit, which
reduces the slope in $f_\textrm{\sc pdr}$.

In contrast, for Ca isotopes,
the sudden jump at $N=28\to 30$ survives in the present calculation,
mainly because of a large shell gap at $N=28$
which makes the neutron pairing gap vanish ($\Delta_n=0$).
At $N=34$, $f_\textrm{\sc pdr}$ in
the present calculation becomes
smaller than the result of HF+RPA. 
This is due to the pairing effect.
In the HF calculation, the ground state in $^{54}$Ca correspond to
the full occupation of the neutron $p_{1/2}$ orbit.
However, the HF+BCS calculation produces fractional occupation
probability of $p_{1/2}$ (52.4 \%) and $f_{5/2}$ orbits (18.2 \%).

\subsection{Medium-heavy nuclei in the neutron-rich side
(28 $\leq Z \leq$ 50 and $N\geq Z$)}
\label{sec:NND-PDR_H}

\begin{figure}[h]
\includegraphics[height=0.45\textwidth, angle=-90]{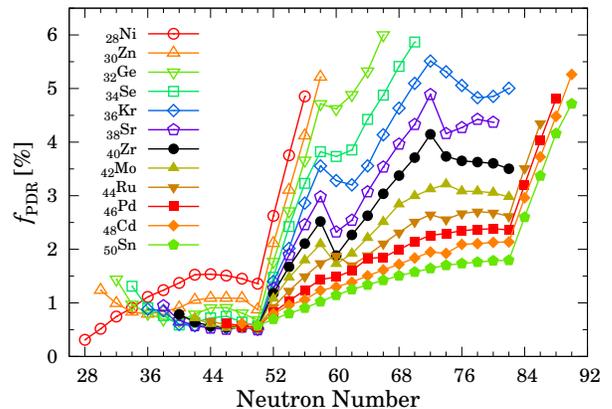}
\caption{The PDR fraction $f_\textrm{\sc pdr}$
as functions of the neutron number for 
even-even isotopes with $Z=28-50$ and $N\geq Z$.
}
\label{fig: H_PDR}
\end{figure}

In Fig. \ref{fig: H_PDR}, we show
the neutron number dependence of
$f_\textrm{\sc pdr}$ for isotopes with $Z=28-50$. 
The characteristic cusps can be seen at $N=$ 50 and 82,
which correspond to the neutron magic numbers.
While the neutrons are filling the $g_{9/2}$ intruder orbit ($40<Z\leq 50$),
the PDR fraction stays approximately constant with respect to the
neutron number.
Especially, the isotopes with $Z=36-50$ (Kr$-$Sn) have roughly identical values
of $f_\textrm{\sc pdr}$ at $N\leq 50$.
Beyond $N=50$, the neutrons start occupying the $d_{5/2}$ orbit,
then, the rapid increase in $f_\textrm{\sc pdr}$ is clearly observed in
Fig. \ref{fig: H_PDR}.
These are universal for all the isotopes shown in Fig. \ref{fig: H_PDR},
although the cusp behavior is weakened by increasing the
proton number.
These are similar to the cusps at $N=14$ and 28 in
lighter neutron-rich isotopes.

In addition, for the isotopes with $Z=32-44$,
the convex cusps also appear at around $N=58-60$
and around $N=72-74$, while 
the concave ones can be seen at $N=60-62$.
These cusps are most prominent around the proton subshell ($Z\approx 40$),
while they become weaker as approaching the magic numbers, $Z\to 28$ and
$Z\to 50$.
This suggests that these may be associated with the ground-state deformation.

Let us briefly comment on the deformation effect.
In Ne and Mg neutron-rich isotopes, the present results suggest
that the onset of deformation in the ground state
increases $f_\textrm{\sc pdr}$.
However, the behavior in the present mass region is more complex. 
$f_\textrm{\sc pdr}$ decrease at the onset of deformation around $N=60$,
then, they decrease again near $N=74$ which corresponds to the disappearance
of the deformation back to the spherical shape. 
In Sec. \ref{sec:Df-PDR}, we discuss the effect of deformation in more
details.

The next jump in $f_\textrm{\sc pdr}$ at $N=82\to 84$ is clearly identified.
This suggests that the definition of the ``low-$\ell$ orbits''
is different between light and heavy systems.
The nuclei around $N=82$ are all calculated to be spherical,
thus it cannot be the effect of deformation.
The single-particle orbit just above the $N=82$ shell gap is
$f_{7/2}$.
In light nuclei, when the Fermi level is located at the $f_{7/2}$
intruder orbits ($20<N\leq 28$), the PDR fraction does not increase
(see Fig.~\ref{fig: L2_PDR}).
The behavior of $f_\textrm{\sc pdr}$ in the heavy isotopes
seems to be very different from that in light systems.
The $f$ orbit may be regarded as the low-$\ell$ orbit for
heavy nuclei with $N>82$.

Finally, the effect of pairing should be noted. 
In Ref.~\cite{INY11}, the similar studies with the HF+RPA were reported
for isotopes with $Z\leq 40$.
The neutron shell effect on $f_\textrm{\sc pdr}$ is qualitatively identical.
However, the HF calculation for heavy isotopes shows peculiar 
changes in the ground-state deformation
from one nucleus to the next, 
which leads to irregular behaviors in $f_\textrm{\sc pdr}$ in
the region of $N>56$.
These irregular behaviors in Ref.~\cite{INY11} are hindered
in the present study.
This is due to the pairing correlation which
produces the fractional occupation probabilities, to
suppress the sudden changes in deformation from nucleus to nucleus.
Now, some systematic trends in the region $56<N<82$ can be observed
in Fig.~\ref{fig: H_PDR}.
For instance, as increasing the proton number from $Z=28$, 
the kink behavior around $N=60$ becomes sharper
toward $Z=38$ (Sr) and 40 (Zr),
then, beyond $Z=40$, it becomes weaker
and disappears near $Z=50$ (Cd and Sn).
The similar systematic behavior can be also observed for
kinks around $N=72$.


\subsection{Neutron skin thickness and PDR fraction}

The classical picture of the PDR is
a vibration of neutron-skin against the core part, from which,
the correlation between skin-thickness and PDR is expected.
In this section, we perform a systematic investigation and
present the correlation between $f_\textrm{\sc pdr}$ and
the skin thickness $\Delta r_\textrm{rms}$ for many isotopes.
The neutron skin thickness is defined by
the difference in root-mean-square radius of neutrons and protons, 
$\Delta r_\textrm{rms}\equiv \sqrt{\langle r^2 \rangle_n} - \sqrt{\langle r^2 \rangle_p}$.

\begin{figure}[h]
  \begin{center}
\includegraphics[height=0.4\textwidth,angle=-90]{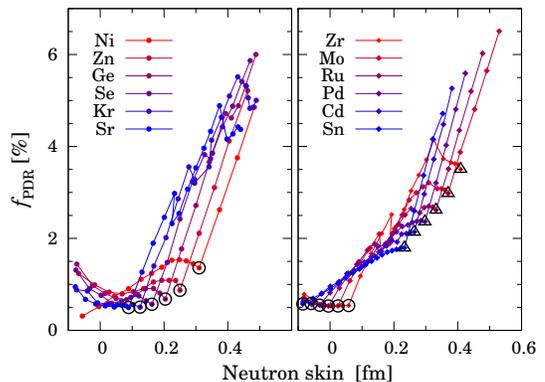}
\caption{The PDR fractions as functions of
the neutron skin thickness for even-even isotopes with $Z=28-50$. 
The open circles (triangles) indicate those with $N=50$ (82).}
	\label{fig: S-P1}
  \end{center}
\end{figure}

Figure \ref{fig: S-P1} shows
$f_\textrm{\sc pdr}$ as a function of the neutron skin thickness.
For Ge, Se, Kr, Sr and Zr isotopes, the similar investigation was performed
with the HF+RPA \cite{INY11}.
A consistent behavior with Ref.~\cite{INY11} is confirmed
in the left panel of Fig.\ref{fig: S-P1}.
There is a linear correlation between $f_\textrm{\sc pdr}$ and $\Delta r_\textrm{rms}$
for $N=50-58$ with a roughly identical slope.
In fact, the slope, $df_\textrm{\sc pdr}/d\Delta r_\textrm{rms}$,
gradually decreases as the proton number increases. 
This can be clearly seen in the right panel of Fig.~\ref{fig: S-P1}
for isotopes with $Z=40-50$. 
As the proton number increases, 
the neutron Fermi level is more deeply bound,
and the slope at $N=50-58$ decreases from Mo to Sn.
The slope $df_\textrm{\sc pdr}/d\Delta r_\textrm{rms}$ 
in Sn isotopes is roughly half of those shown in the left panel
of Fig.\ref{fig: S-P1}. 
Beyond $N=82$,  $df_\textrm{\sc pdr}/d\Delta r_\textrm{rms}$ again increase and
stay almost constant (a linear correlation).
The observed slopes at $N>82$ is almost identical to those
in the left panel with $Z=28-38$ at $N=50-58$. 

The properties of the correlation between the PDR strength and
the neutron skin thickness turns out to be rather complex,
depending on both the proton and the neutron numbers.
Nevertheless, in the present calculation,
we observe a universal behavior which is the maximum gradient for Ni isotopes, 
$(df_\textrm{\sc pdr}/d\Delta r_\textrm{rms})_\textrm{max}\approx 0.26$ fm$^{-1}$.
The maximum values appears at $N>50$ and $N>82$
in which the neutron Fermi levels are located at weakly-bound
$d_{5/2}$ and $f_{7/2}$ orbits, respectively.
It should be noted that the similar maximum value of
$(df_\textrm{\sc pdr}/d\Delta r_\textrm{rms})_\textrm{max}\approx 0.2$ fm$^{-1}$ 
was previously observed in the HF+RPA calculation
in lighter mass region at $N>28$
in which the neutrons are filling weakly-bound $p$ orbits \cite{INY11}.

\subsection{Analysis with the SkI3 energy functional}
\label{sec:cSkI3}

The PDR properties predicted by the relativistic mean-field (RMF) theories
are often quantitatively different from those obtained with the non-relativistic
Skyrme energy functionals,
although they show qualitative agreements \cite{PVKC07}.
For instance, in the present calculation with the SkM$^*$ functional, 
the PDR strength
$f_\textrm{\sc pdr}$ for $^{132}$Sn is less than 2 \%,
while it is more than 4 \% in the relativistic calculation
with the DD-ME2 \cite{PVKC07}, FSUGold, and NL3 \cite{Piekarewicz06}. 
The experimental data suggested the value of about 4 \% \cite{Adrich}. 
To investigate the origin of the difference, 
we adopt the SkI3 Skyrme parameter set which has the same
density dependence of the spin-orbit form factor as that of the RMF \cite{RF95}.

\begin{center}
\begin{figure*}[t]
\includegraphics[keepaspectratio,width=50mm, angle=-90]{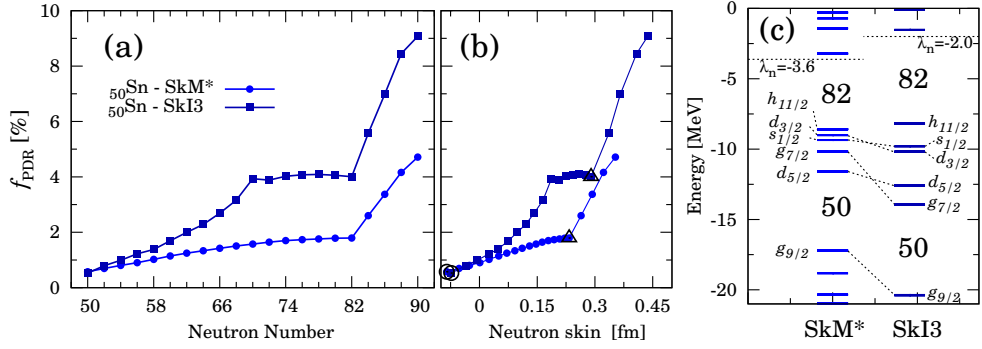}
\caption{(a) The PDR fraction of Sn isotopes as a function of neutron number,
calculated with the SkM$^*$ energy functional (circles) 
and with SkI3 (squares).
(b) The same as (a) but as a function of the neutron skin thickness.
Open circles (triangles) indicate $N=50$ (82).
(c) Neutron single-particle levels around chemical potential $\lambda_n$ 
for $^{134}$Sn, calculated with SkM$^*$ and SkI3 functionals.
}
\label{fig: ID_P-S}
\end{figure*}
\end{center}

Figure \ref{fig: ID_P-S} (a) shows the neutron number dependence
of $f_\textrm{\sc pdr}$ in Sn isotopes,
which are obtained by Cb-TDHFB calculation with SkM$^*$ (filled circle)
and SkI3 (filled square).
We can see a clear difference between them.
The result of SkI3 show rapid increase of $f_\textrm{\sc pdr}$ as a function
of neutron number in the regions of $58<N<70$ and $N>82$.
It shows constant values of $f_\textrm{\sc pdr}\approx 4$ \%
in $70\leq N\leq 82$, corresponding to the neutron 
Fermi level in the intruder $h_{11/2}$ orbit.
These behaviors produce prominent kinks at $N=70$ and 82
in Fig.~\ref{fig: ID_P-S} (a).
This is consistent with the RMF results of Refs.~\cite{Piekarewicz06,PVKC07}. 

The single-particle levels of $^{134}$Sn obtained with the SkM$^*$
and the SkI3 parameter sets are shown in Fig.~\ref{fig: ID_P-S} (c).
The level spacings in the region of $50<N<82$
are much larger in SkI3 than in SkM$^*$. 
Note that the effective mass ($m^\ast\!/m$) in SkI3 is 0.58 which is
significantly smaller than 0.79 in SkM$^*$. 
In addition, the level ordering of $g_{7/2}$ and $d_{5/2}$
are different.
These properties lead to prominent subshell effect on the PDR with SkI3.
Namely, after the full occupation of $g_{7/2}$ orbit at $N=58$,
while the neutrons start to fill the low-$\ell$ ($d$ and $s$) orbits,
$f_\textrm{\sc pdr}$ shows a rapid increase.
Then, beyond $N=70$, $f_\textrm{\sc pdr}$ stops increasing,
since the neutrons now fill the intruder orbit $h_{11/2}$.
These effects are much weaker in SkM$^*$ because the single-particle
levels are much denser and the subshell effects are smeared out.

In Fig.~\ref{fig: ID_P-S} (b), we show $f_\textrm{\sc pdr}$
as a function of the neutron skin-thickness.
It shows a clear discrepancy between the results with SkM$^*$ and SkI3,
which is associated with the different subshell effects.
Apparently, the PDR strength is not uniquely determined by the
neutron skin thickness, but affected by the choice of the energy
functional and associated single-particle structure.
In contrast, the rapid increase of $f_\textrm{\sc pdr}$ at $N>82$
is a universal feature.
The slope $df_\textrm{\sc pdr}/d\Delta r_\textrm{rms}$ is also similar
to each other,
although the slope in SkI3,
$df_\textrm{\sc pdr}/d\Delta r_\textrm{rms}\approx 0.37$ fm$^{-1}$,
is slightly steeper than that in SkM$^*$.

In summary, the PDR strength significantly depends on
the single-particle shell structure, thus has a functional dependence.
In fact, even in the relativistic calculations,
the ones adopting the point-coupling energy functionals
show results very similar to the present calculation with SkM$^*$
\cite{DR11}.
This strong shell effect of the PDR may be useful for investigating
the neutron single-particle levels in unstable nuclei.

\section{Deformation effects}
\label{sec:Df-PDR}

It is well-known that the nuclear deformation in the ground state
significantly affects the shape of the GDR peak \cite{BF75,BM75,HW01,YN11}.
For a well-deformed nucleus, the GDR peak is split into two peaks,
which can be understood as a simple geometrical effect.
Namely, there are different frequencies associated with long and short
principal axes of deformation.
On the other hand, for the low-lying $E1$ mode, the geometrical effect
is not prominent.
Ref.~\cite{PKR09} has reported that the deformation hinders the PDR
strength for the prolate neutron-rich Sn isotopes.
They also found that the $K=0$ component of calculated low-lying $E1$
strength is larger than $K=1$ component, which was interpreted 
by the effect of anisotropic neutron skins.
Namely, the neutron skin thickness in the $z$ ($K=0$) direction is expected to
be thicker than the $x-y$ ($K=1$) directions. 
In this section, we investigate this deformation effect for selected
isotopes; strongly deformed (Sr and Zr) and
weakly deformed nuclei (Pd and Cd).

\begin{figure}[tbh]
   \begin{center}
    \includegraphics[keepaspectratio,width=50mm, angle=-90]{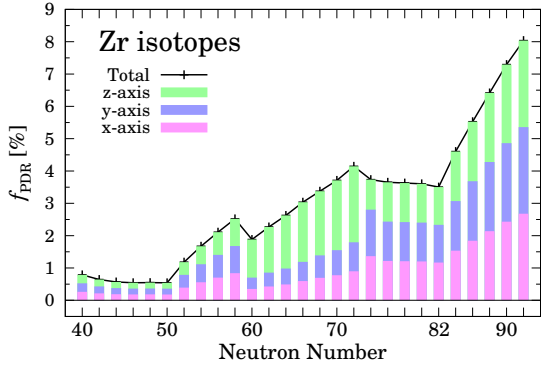}
	\caption{The PDR strength in Zr isotopes decomposed into 
$x$, $y$, and $z$ components which are denoted by red, blue, and green bars.
In spherical nuclei, the three components are equal to each other.
In this figure, we include isotopes with the neutron chemical potentials
smaller than 2 MeV ($N>82$).
}
\label{fig: Zr_10}
  \end{center}
\end{figure}

\subsection{Deformation and PDR strength}

In Fig.~\ref{fig: Zr_10}, $f_\textrm{\sc pdr}$ 
are shown for Zr isotopes.
In the present calculation, Zr isotopes with $N=60-72$ are deformed with
a prolate shape and the one with $N=74$ has a small triaxial shape.
The other nuclei are calculated to be spherical in the ground state.
Here, we can clearly identify the deformation effects.
The total PDR strength suddenly reduces at $N=58\rightarrow 60$
at the onset of deformation ($\beta_2\approx 0.37$).
This is consistent with Ref.~\cite{PKR09} that the deformation
hinders the PDR strength.
However, at $N=72\rightarrow 76$ where the deformation reduces and
disappears, the PDR strength decreases again.
Thus, the deformation does not always hinder the PDR strength.

\begin{figure}[h]
  \begin{center}
\includegraphics[keepaspectratio,width=50mm, angle=-90]{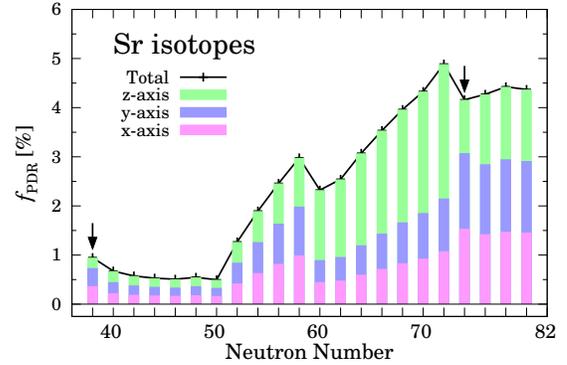}
\caption{Same as Fig.\ref{fig: Zr_10}, but for Sr isotopes. 
The arrows indicate the oblate-shape nuclei.}
\label{fig: Sr_10}
   \end{center}
\end{figure}

Very similar behaviors can be found for Sr isotopes as well,
in Fig.~\ref{fig: Sr_10}.
In Sr isotopes, the ground state has a prolate shape with
$\beta_2\approx 0.37$ for $N=60-72$, and an oblate shape at $N=38$ and $74$.
When the prolate deformation develops in the ground state,
the PDR strength reduces.
The largest PDR strength is seen in $N=72$, then it drops at $N>72$
as the ground state is going back to the spherical shape.

\begin{figure}
   \begin{center}
\includegraphics[keepaspectratio,width=50mm, angle=-90]{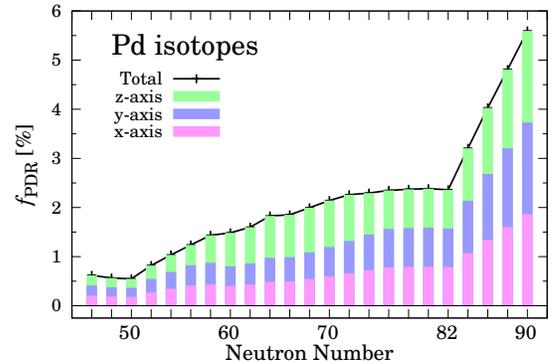}
\caption{Same as Fig.\ref{fig: Zr_10}, but for Pd isotopes.}
\label{fig: Pd_10}
  \end{center}
\end{figure}

In weakly deformed isotopes, the deformation effect on the PDR is
much milder.
In Fig. \ref{fig: Pd_10}, we show Pd isotopes. 
The ground state of Pd isotopes is deformed with a prolate shape
at $N=58-74$, however, their deformation $\beta_2=0.1-0.2$
is significantly smaller than Zr isotopes.
The PDR strength gradually increases toward $N=82$, then, jump up
beyond that.
In Cd isotopes, we observe almost identical behaviors to Pd.


\begin{figure*}[t]
 \begin{minipage}{0.45\hsize}
  \begin{center}
\includegraphics[keepaspectratio,width=50mm, angle=-90]{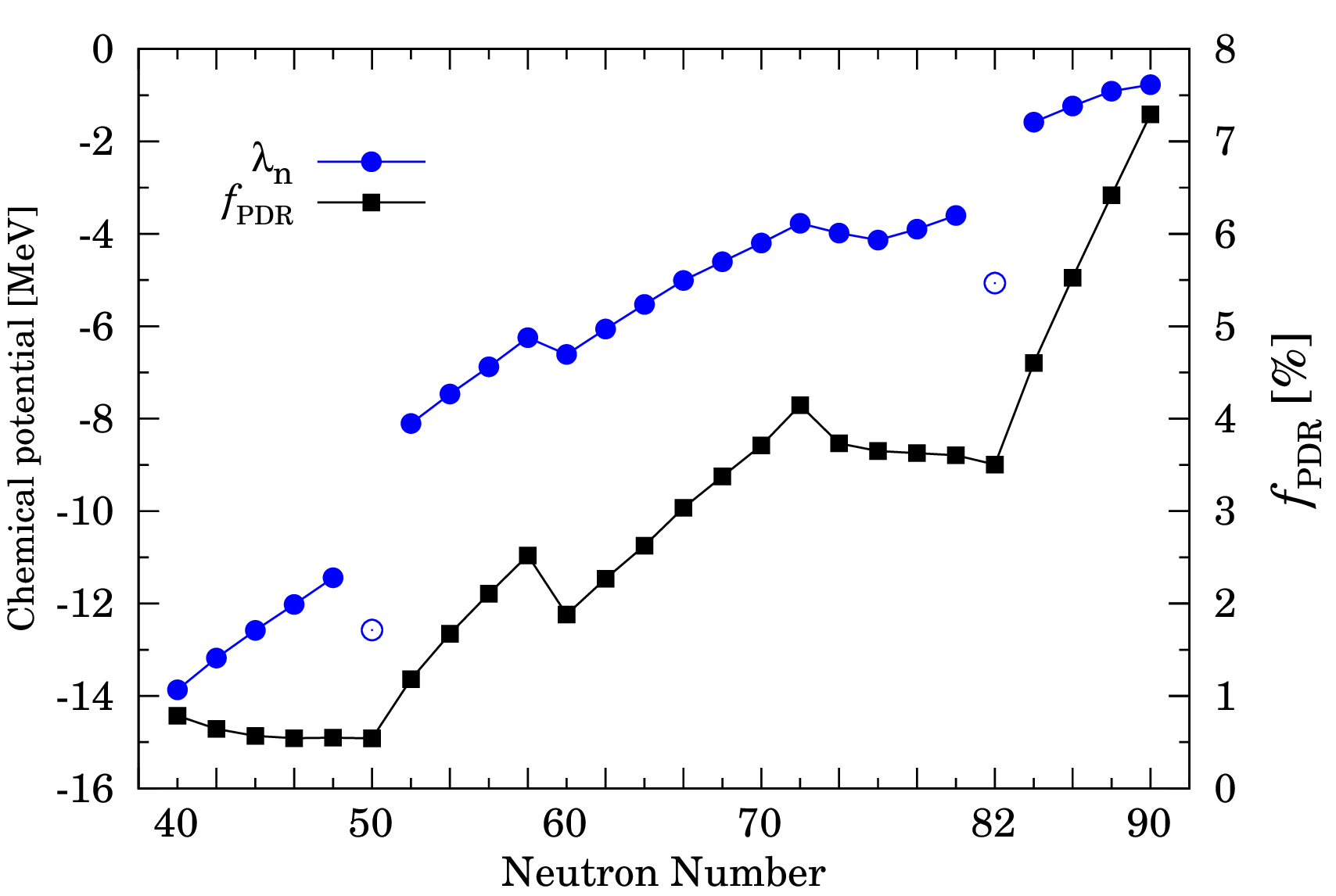}
\caption{
Neutron chemical potentials for Zr isotopes as a
function of neutron number are shown by circles.
At N = 50 and 82, the neutron pairing gaps vanish,
for which the highest-occupied
single-particle levels are shown by open circles.
The PDR fraction $f_\textrm{\sc pdr}$ are also shown
by squares and their scale is given in the right axis.
}
\label{fig: Cp-Zr}
   \end{center}
 \end{minipage} \hspace{5mm}
 \begin{minipage}{0.45\hsize}
   \begin{center}
\includegraphics[keepaspectratio,width=50mm, angle=-90]{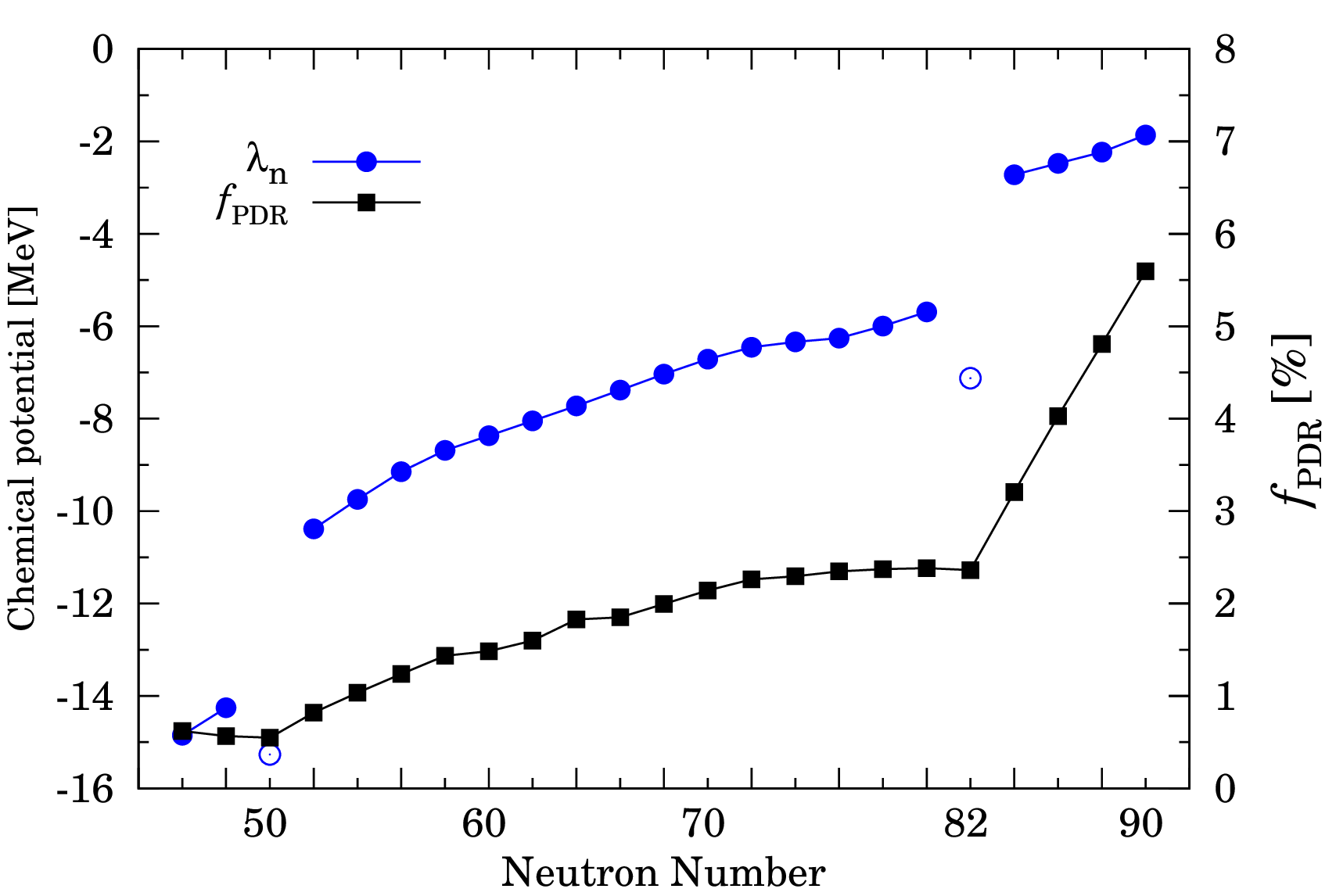}
\caption{Same as Fig.\ref{fig: Cp-Zr}, but for Pd isotopes.}
\label{fig: Cp-Pd}
  \end{center}
 \end{minipage}
\end{figure*}

We have found that the isotopic dependence of the PDR strength
is consistent with the behavior
of the neutron chemical potentials.
In Figs.~\ref{fig: Cp-Zr} and \ref{fig: Cp-Pd},
the neutron chemical potentials $-\lambda_n$ are shown for Zr and Pd
isotopes, respectively.
In Zr isotopes, the sudden change of the deformation, in fact, causes
the change in $\lambda_n$ as well.
The chemical potential $|\lambda_n|$ monotonically decreases as
increasing the neutron number, except for two places.
These exceptional increase of $|\lambda_n|$
takes place at $N=58\rightarrow 60$
and at $N=72\rightarrow 76$.
These neutron numbers perfectly agree with those showing
the drops of the PDR strength in neutron-rich Zr isotopes.
In contrast, the $|\lambda_n|$ in Pd isotopes show a monotonic decrease,
leading to a monotonic increase in $f_\textrm{\sc pdr}$.
Therefore, it seems that the deformation affects the PDR strength
through changing the neutron chemical potential.

\subsection{The $K=0$ dominance in PDR strength}

In GDR at high frequency $E1$ mode, the deformation splitting was
systematically observed in deformed nuclei.
The $K=0$ component is lower in energy than $K=1$ component,
leading to a double-peak structure in the prolate nuclei.
In oblate nuclei, the opposite situation is expected, so that
the $K=0$ is higher than $K=1$. 
In Figs.~\ref{fig: Zr_10}, \ref{fig: Sr_10}, and \ref{fig: Pd_10},
the PDR strengths are decomposed into $x$, $y$, and $z$ components. 
Here, $z$ direction is chosen as the symmetry axis of the deformation.
The calculation indicates that, 
in the prolate nuclei, 
the $K=0$ ($z$) PDR components are larger than the $K=1$ components.
This is consistent with Ref.~\cite{PKR09}, in which it was interpreted
to be due to the difference in the neutron skin thickness with respect
to the $z$ and $x-y$ directions.
In Ref.~\cite{PKR09}, for very neutron-rich Sn isotopes with $N=92-112$,
the prolate deformation of the neutron density was calculated to be larger than
that of protons ($(\beta_2)_n>(\beta_2)_p$).
This leads to larger neutron skin in $z$ direction of the symmetry axis
than in the perpendicular directions ($x-y$).
However, it turns out that, in the present calculation, 
the situation in Sr and Zr isotopes is very different from Sn isotopes in Ref.~\cite{PKR09}.

\begin{figure*}[t]
 \begin{minipage}{0.45\hsize}
  \begin{center}
    \includegraphics[keepaspectratio,width=50mm, angle=-90]{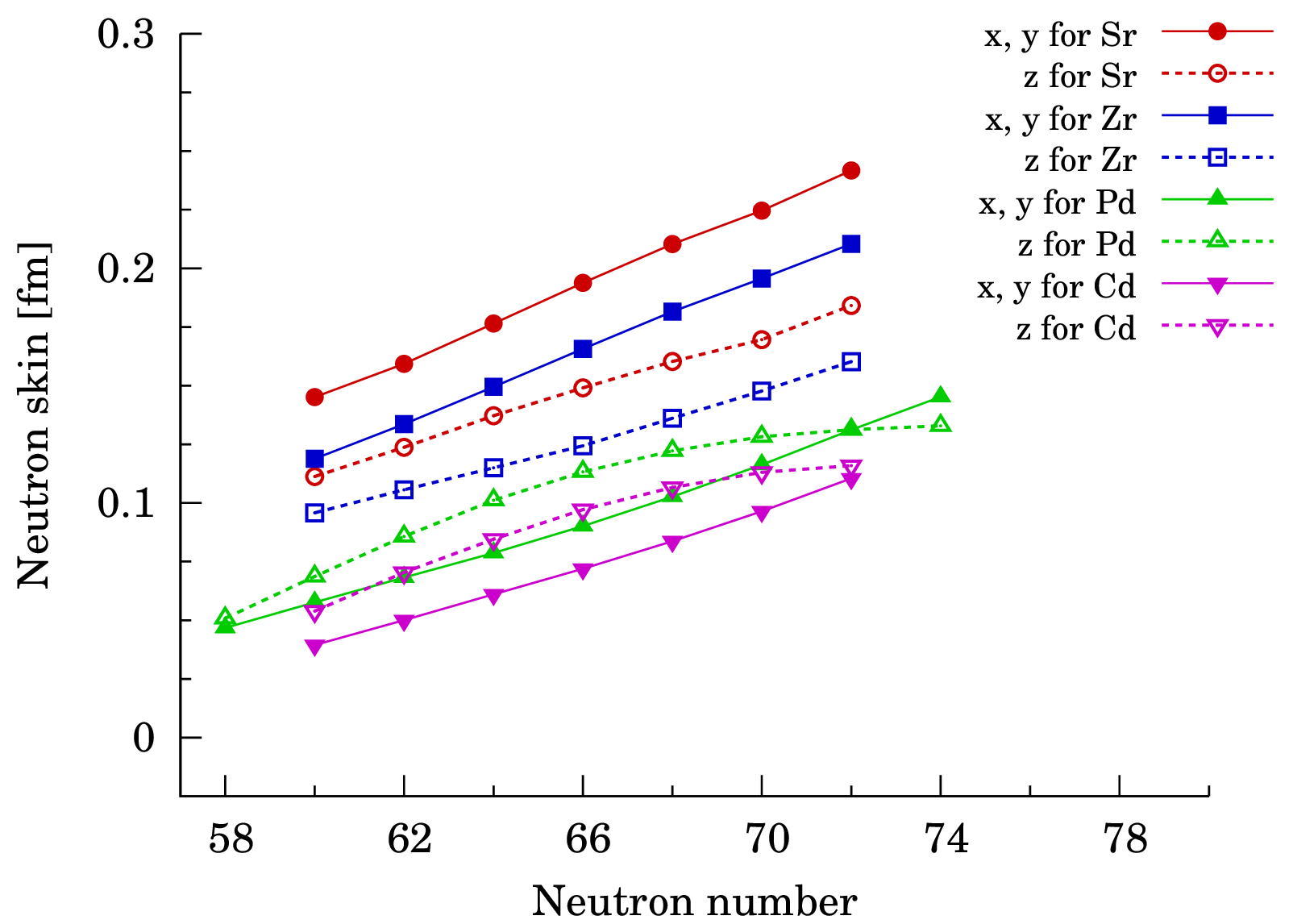}
	\caption{Neutron skin thickness in different directions for prolate deformed Sr, Zr, Pd and Cd isotopes.
                 The values along $z$ and the perpendicular directions are shown by empty and filled symbols, respectively.
}
	\label{fig: pro-skin}
   \end{center}
 \end{minipage} \hspace{5mm}
 \begin{minipage}{0.45\hsize}
   \begin{center}
   \includegraphics[keepaspectratio,width=50mm, angle=-90]{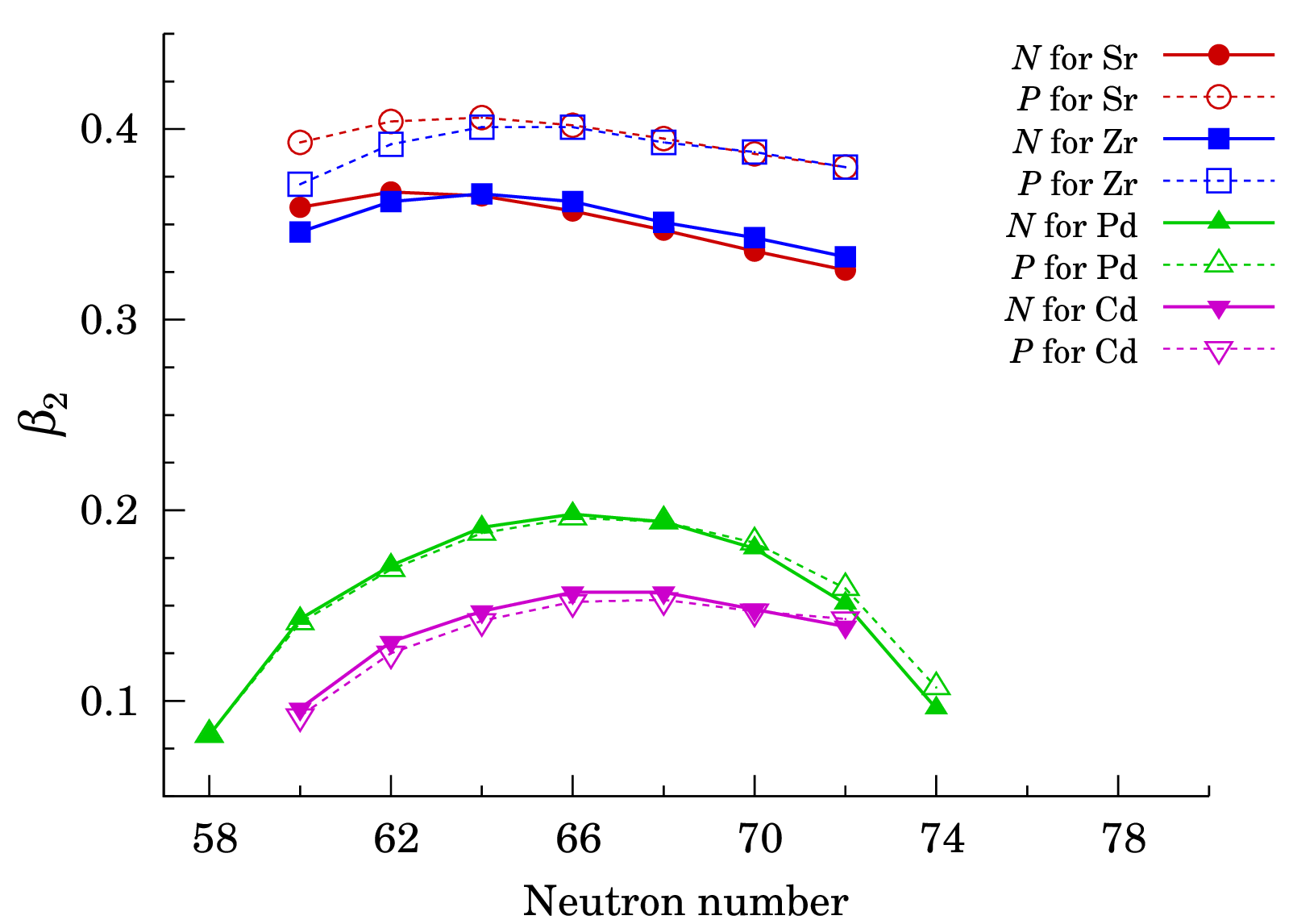}
   \caption{Neutron and proton quadrupole deformations, $(\beta_2)_n$ and $(\beta_2)_p$, 
are respectively shown by filled and open symbols,
 for prolate deformed Sr, Zr, Pd and Cd isotopes.}
	\label{fig: beta-pro}
  \end{center}
 \end{minipage}
\end{figure*}

Figure \ref{fig: pro-skin} shows the neutron skin thickness with respect
to $z$ and $x-y$ directions for prolate deformed 
Sr, Zr, Pd and Cd isotopes.
These are defined by the root-mean-square radius of the $z$ direction,
$\sqrt{\langle z^2 \rangle_n} - \sqrt{\langle z^2 \rangle_p}$, 
shown by dashed lines,
and the same for $y$ and $z$-directions shown by solid lines. 
It turns out that,
in Sr and Zr isotopes, the neutron skin is thicker in the $x-y$ plane
than in the $z$ direction,
while the situation is opposite in Pd and Cd isotopes.

In Fig.~\ref{fig: beta-pro}, we show the calculated quadrupole deformation
$\beta_2$ for protons and neutrons separately.
In Pd and Cd isotopes, the deformation of the proton density distribution
is identical to that of neutrons.
However, for Sr and Zr isotopes, the deformation is actually larger for
protons than for neutrons.
Thus, the skin thickness enhances in the perpendicular directions ($x-y$)
as is shown in Fig.~\ref{fig: pro-skin}.
If the neutron skin thickness is directly related to the PDR strength,
we could expect the $K=1$ dominance, rather than the $K=0$ dominance.
This is opposite to what we have observed
in the present calculation (Figs.~\ref{fig: Zr_10} and \ref{fig: Sr_10}).
The anisotropy of the the neutron skin thickness
is unlikely to be the origin of the $K=0$ dominance in the PDR strength.

Here, we give another possible mechanism of the $K=0$ dominance
in prolate deformed nuclei, which is nothing to do with the neutron skin
nor the weak binding nature of the valence neutrons.
Let us discuss single-particle matrix elements of the dipole operators
in the harmonic oscillator potential model.
The single-particle states are labeled by the oscillator quanta
$(n_x,n_y,n_z)$ which correspond to the single-particle energies,
$\epsilon_{n_x n_y n_z}=\omega_x n_x + \omega_y n_y + \omega_z n_z + 3/2$.
The well-known geometric effect, $\omega_z <\omega_x$ for a prolate nucleus,
lowers the excitation energy of $K=0$ modes ($\omega_z$)
compared to $K=1$ excitations ($\omega_x=\omega_y$).
The $K=0$ dominance of the PDR strength can be partially due to this effect,
but this is not the only cause.
In fact, the deformation in the ground state naturally leads to
the $K=0$ enhancement of the dipole matrix elements.
It is easy to see that the ratios of available dipole matrix elements
between $z$ and $x$ ($y$) directions are given by
\begin{equation}
\left|
\frac{\bra{n_x,n_y,n_z+1} z \ket{n_x,n_y,n_z}}
{\bra{n_x+1,n_y,n_z} x \ket{n_x,n_y,n_z}} \right|^2
 = \frac{n_z+1}{n_x+1}\frac{\omega_x}{\omega_z} .
\end{equation}
For the ratio $|\langle z \rangle/\langle y \rangle|^2$,
we have the same equation but simply replacing $x$ by $y$.
The $K=0$ strength is enhanced by the fact that
$\omega_x/\omega_z= \omega_y/\omega_z > 1$ for prolate nuclei.
For strongly deformed Sr and Zr isotopes with $\beta_2\approx 0.37$,
this factor amounts to about 1.5.
In addition, for occupied (hole) orbitals,
the average value of $n_z$ is larger than $n_x$.
This also contributes to the $K=0$ enhancement.

\section{Systematic behaviors in PDR strengths around $N$=82}
\label{sec:A82-PDR} 

It is a well-known fact that most of the $E1$ strengths in nuclei are carried
by the GDR at high frequency.
Therefore, the low-energy $E1$ strengths are strongly hindered,
typically with a factor of $10^{-5}$ in heavy nuclei.
In the terminology of the present time-dependent calculation,
the time-dependent residual induced field,
$\delta h(t)=\delta h/\delta\rho \cdot \delta \rho(t)$,
significantly hinders the low-lying single-particle $E1$ strength.
However, in the present study, we show that
the PDR peaks in the neutron-rich nuclei at $N>82$ demonstrate
some exceptional cases.
Namely, low-energy $E1$ strengths appearing around $E=5$ MeV
are completely decoupled from the GDR.
Both $E1$ strengths and peak positions remain same as those of
the corresponding unperturbed (single-particle) peaks.
In this section, we demonstrate that these striking behaviors can be
universally observed in the PDR peaks
in heavy neutron-rich isotopes beyond $N=82$.

\subsection{Choice of smoothing function}

So far, we have used the exponential decay function,
$w(t)=e^{-\Gamma t/2}$,
as a smoothing function in Eq. (\ref{F(E)}).
This produces the strength function $S(E;E1)$ in which the delta functions
in Eq. (\ref{S(E1)}) are smeared by 
the Lorentzian function of a width $\Gamma$.
In order to identify detailed structure of the low-energy $E1$ peaks,
the width of $\Gamma=1$ MeV is too large.
Each peak has a rather long tail with an asymptotic behavior of $1/|E-E_n|^2$,
which makes individual peaks strongly overlap with each other.

Thus, in this section, we use a different smoothing function $w(t)$
in Eq. (\ref{F(E)}),
\begin{equation}
w(t)=1-3(t/T)^2+2(t/T)^3 .
\label{3-ji-shiki}
\end{equation}
This function has properties, $w(0)=1$, $w(T)=0$, and $w'(0)=w'(T)=0$,
which leads to good conservation of the EWSR value in practice.
Instead of the Lorentzian function, a peak at $E=E_n$ is smeared as
\begin{eqnarray}
{\cal F}(E\approx E_n) &\sim &
\textrm{Re} \int_0^T e^{i(E-E_n)t} w(t) dt \nonumber \\
&=& \frac{6T}{x^3}
\left\{
-\sin x + \frac{2(1-\cos x)}{x}
\right\} ,
\end{eqnarray}
where $x=(E - E_n) T$.
This smearing function quickly disappears when the energy $E$ is away
from the peak position $E_n$.
In fact, the function becomes zero at $|E-E_n|=2\pi/T$ and
negligibly small at $|E-E_n|>2\pi/T$.

\subsection{PDR peaks with and without residual fields}

\begin{figure*}[t]
\includegraphics[keepaspectratio,width=45mm, angle=-90]{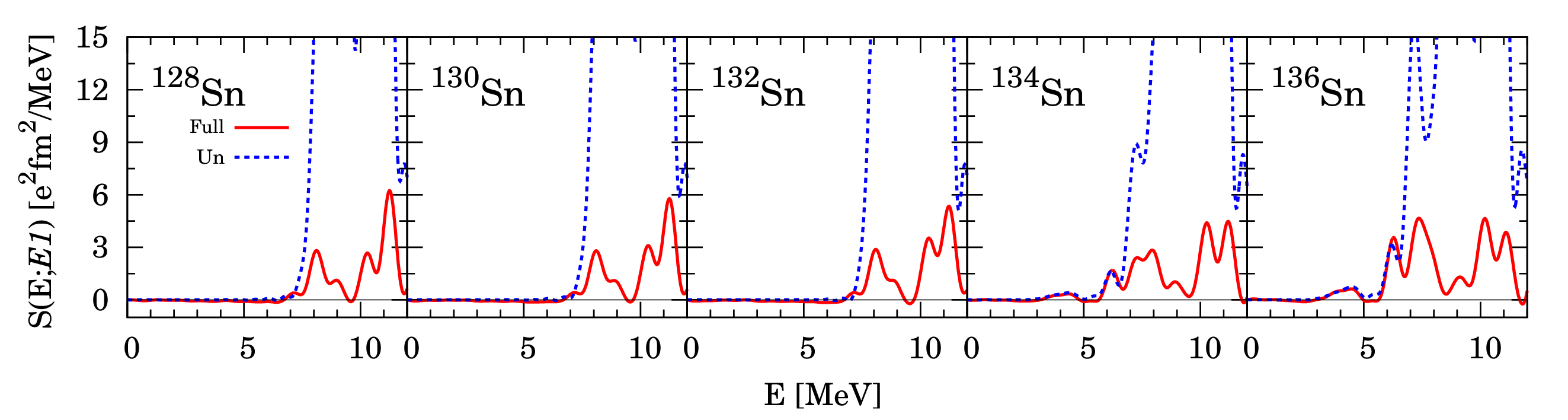}
\caption{$E1$ strength functions below 12 MeV for Sn isotopes with $N=78-86$,
calculated with the SkM$^*$ parameter set.
The solid and dashed lines show full and unperturbed strength distributions.
For the function $w(t)$, Eq. (\ref{3-ji-shiki}) is used.}
\label{fig: Sn_SkM}
\includegraphics[keepaspectratio,width=45mm, angle=-90]{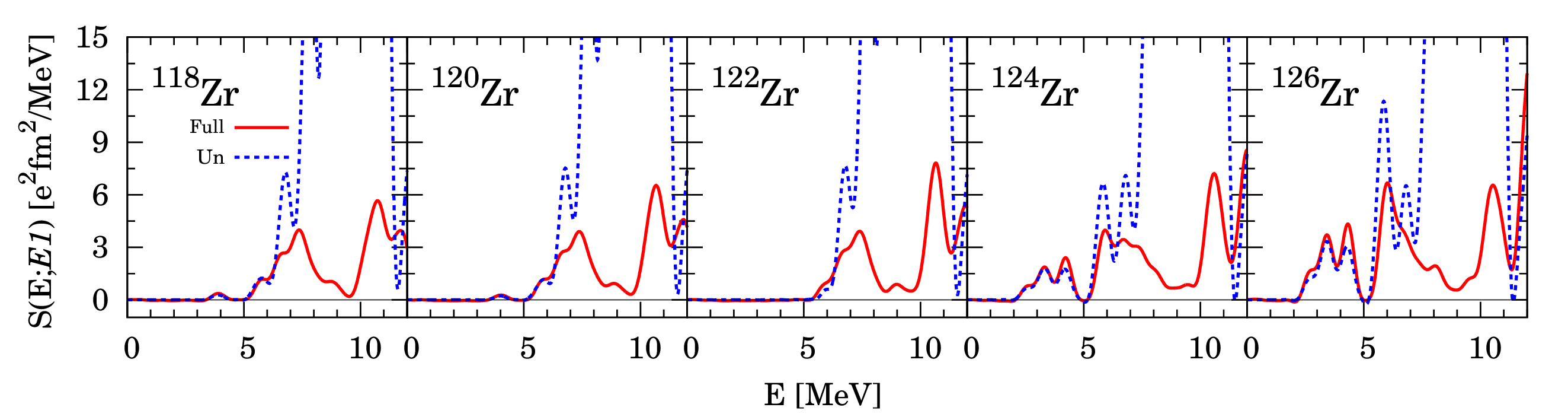}
\caption{Same as Fig. \ref{fig: Sn_SkM}, but for Zr isotopes.}
\label{fig: Zr_SkM}
\end{figure*}

In order to examine the low-energy $E1$ peaks, we compare the Cb-TDHFB
results with the unperturbed results.
The unperturbed results can be obtained by neglecting all the residual
induced fields.
Namely, during the time evolution, we simply
ignore the time-dependence of mean fields, $h(t)$ and $\Delta(t)$,
in Eq. (\ref{Cb-TDHFB}), 
by keeping all the fields same as those in the ground state,
$h(t)=h_0$ and $\Delta(t)=\Delta_0$.
The calculated $E1$ strength functions at $E<12$ MeV are shown in
Fig.~\ref{fig: Sn_SkM} for Sn isotopes.
For Sn isotopes with $N\leq 82$ ($^{128-132}$Sn),
the $E1$ strengths appear in the energy region of $E>7$ MeV.
They are almost identical to each other,
and are significantly hindered from the unperturbed strengths.
This PDR structure at $E=7-12$ MeV remains similar even for isotopes
beyond $N=82$ ($^{134-136}$Sn).
However, there appear other low-energy peaks around $E=5$ MeV.
In Fig.~\ref{fig: Sn_SkM}, the dashed and solid lines coincide
with each other around $E=5$ MeV.
These low-energy dipole peaks appearing in Sn isotopes
with $N>82$ are not hindered by the residual effect.
In other words,
these dipole modes of excitation have pure single-particle nature,
being decoupled from the GDR.

The similar feature can be observed more clearly in Zr isotopes
in Fig.~\ref{fig: Zr_SkM}.
The PDR peak around $E=7$ MeV exists for all the isotopes
in Fig.~\ref{fig: Zr_SkM},
however, in those with $N> 82$, other low-energy peaks emerge
around $E=2-5$ MeV.
These new peaks at $N>82$ show no hindrance by the coupling with the GDR.

\begin{figure*}[t]
\includegraphics[keepaspectratio,width=50mm, angle=-90]{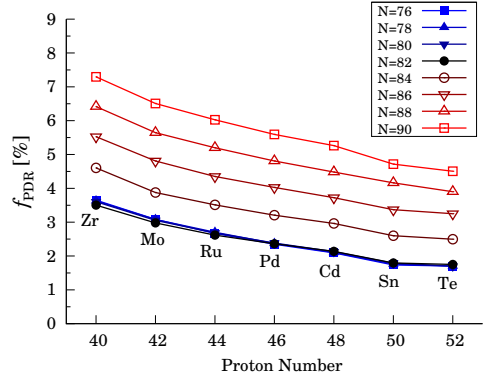}
\caption{Proton number dependence of the PDR fraction for Zr, Mo, Ru, Pd, Cd, Sn, 
and Te isotopes around $N=82$.}
\label{fig: A82-PDR}
\end{figure*} 

In Fig.\ref{fig: A82-PDR},
we summarize $f_\textrm{\sc pdr}$ of Zr, Mo, Ru, Pd, Cd and Sn isotopes
around $N=82$. 
In each isotopic chain, we find universal behaviors of $f_\textrm{\sc pdr}$:
It stays constant for $76\leq N\leq 82$, then, jumps up beyond $N=82$.
This is due to the fact that two kinds of low-lying dipole modes exist in these isotopes.
One of these appears at $E>7$ MeV whose $E1$ strength is strongly hindered by the residual effect. 
Most probably, it may have a collective isoscalar character\footnote{
Unfortunately, it is difficult to perform the accurate real-time
calculation for the isoscalar dipole modes, because the instantaneous
perturbation $V_\textrm{ext}(t)$ inevitably excites the center-of-mass motion.
The large amplitude center-of-mass motion leads to a significant error
in the Fourier transformation of Eq.(\ref{F(E)}).}\cite{RPBMC12}.
The other one appears at even lower energy of $E=2-7$ MeV.
The $E1$ strength associated with these peaks shows no hindrance,
suggesting that these $E1$ modes are well decoupled from the GDR.

\begin{figure*}[t]
\includegraphics[keepaspectratio,width=45mm, angle=-90]{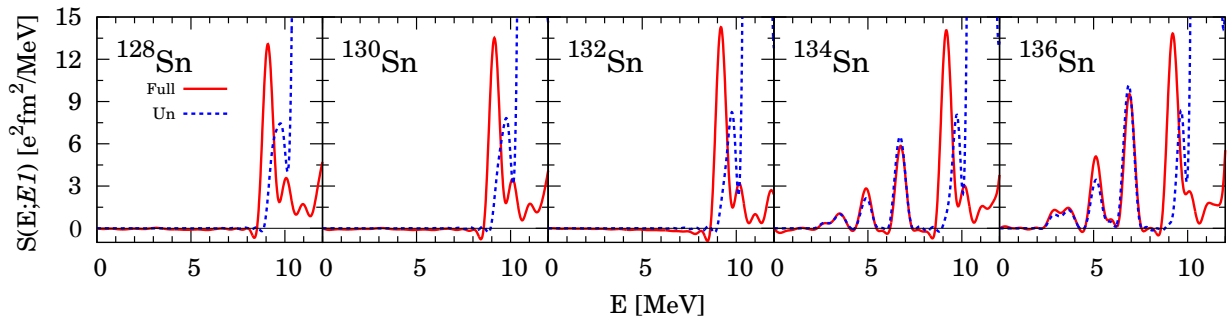}
\caption{Same as Fig. \ref{fig: Sn_SkM}, but with the SkI3 parameter set.}
\label{fig: Sn_SkI3}
\end{figure*}

Finally, in Fig.~\ref{fig: Sn_SkI3},
we show the result of the calculation using the SkI3 functional.
There is a prominent PDR peak at $E\approx 9$ MeV for all Sn isotopes
for $N=78-86$.
In fact, the $E1$ strength of this peak is enhanced by the residual effect.
It means that there is a collective nature of the isovector ($E1$) response.
This is unusual but explains the property discussed in Sec.~\ref{sec:cSkI3}
that the PDR strength in SkI3 is significantly
larger than that in SkM$^*$.
Beyond $N=82$, again we can find the PDR peaks at $E=2-7$ MeV,
which show no hindrance by the residual interactions.

\section{Conclusion}
\label{sec:summary}

We have performed systematic investigation of the $E1$ modes of excitations
in terms of the linear response calculation using the
real-time Cb-TDHFB method in the 3D-coordinate space representation. 
In particular, properties of the PDRs
and their relations to neutron number, neutron skin-thickness, deformation, and chemical potential
have been systematically investigated. 
The present calculation takes into account effects of any kind of deformation
and the static and dynamical pairing correlation in the BCS-like approximation.
We use the Skyrme energy functional of SkM$^*$ for the present study.
For Sn isotopes, we also adopt the SkI3 parameter set 
to discuss the origin of known difference between non-relativistic and relativistic calculations.

We have confirmed a strong neutron shell effect.
Especially, we see magic neutron numbers for the sudden enhancement
of the PDR strength, such as
$N$=14 $\rightarrow 16$,
$N$=28 $\rightarrow 30$,
$N$=50 $\rightarrow 52$, and
$N$=82 $\rightarrow 84$.
These characteristic numbers appear in the results both with and without
pairing correlation.
However, the shell effects are somewhat weakened by the pairing effect
for S and Ar isotopes around $N$=28,
which indicates a gradual increase of the PDR fractions.
This is due to the pairing effect of the fractional occupation probability
of single-particle orbits.

The pairing correlation also removes some irregular behaviors in the PDR
strength in the region of $60<N<74$.
Because of this, we can observe a clear indication of the deformation effect.
It is most prominent in Kr ($Z$=36) , Sr ($Z$=38), and Zr ($Z$=40) isotopes.
There are sudden reduction of the PDR strength at the onset and
the offset of deformation.
In strongly deformed nuclei with the prolate shape, such as Sr and Zr isotopes,
the PDR strength is dominated by the $K=0$ component,
while the neutron skins in these isotopes are smaller 
in the direction of the 
symmetry axis than in the perpendicular directions.
Thus, the $K=0$ dominance is not associated with the skin thickness.
It may be explained, at least partially,
by difference in the single-particle dipole strengths
in the deformed mean-field potential.

We have closely examined the characters of the PDR peaks appearing
around $N$=82.
It turns out that there are two kinds of peaks.
The $E1$ strengths associated with peaks located at $E>$7 MeV are
very different from the unperturbed strengths.
Generally speaking,
they are usually hindered from the unperturbed 
single-particle strengths. 
On the other hand, the peaks appearing at $E<$5 MeV beyond $N$=82
are identical to those in the unperturbed calculation.
This suggests the pure single-particle character
and the decoupling of the low-energy dipole modes from the GDR.
This decoupling phenomenon beyond $N$=82 
can be observed for many kinds of isotopes around Sn ($Z$=50)
and obtained with different energy functionals as well.

It should be noted that properties of the PDRs are sensitive to
underlying nuclear structure, such as the single-particle level. 
For instance, the single-particle level spacings in SkI3 are generally larger
than those in SkM$^*$, due to the smaller effective mass. 
This leads to significant difference
in the PDR strengths for Sn isotopes between SkM$^*$ and SkI3.
The relativistic mean-field calculation with the NL3 parameter set \cite{Piekarewicz06}
produces the result similar to the non-relativistic calculation with SkI3,
while the point-coupling relativistic model \cite{DR11} produces the one similar
to SkM$^*$. Therefore, 
the different prediction on the PDR properties in the literature
may be mostly attributed to the difference in underlying shell structure.

We have observed some
correlation between the PDR strength and the neutron-skin thickness,
however, its property and quality depend on the mass number,
the neutron number, and
the adopted energy functional.
In contrast, the strong enhancement of the PDR strengths at
specific neutron numbers
($N$ = 14 $\rightarrow 16$,
$N$ = 28 $\rightarrow 30$,
$N$ = 50 $\rightarrow 52$, and
$N$ = 82 $\rightarrow 84$)
and the decoupling phenomenon observed beyond $N=82$
are more robust and universal.

\begin{acknowledgments}
This work was supported by JSPS KAKENHI Grant Numbers
24105008, 24105006, 25287065, 2528706 and by the MEXT SPIRE Field 5.
Computational resources were partially provided by the HPCI Systems
Research Projects (hp120192)
and the RIKEN Integrated Cluster of Clusters (RICC). 
\end{acknowledgments}

\appendix*

\section{Ground-state properties}

In this appendix, the calculated ground-state properties 
are presented for even-even isotopes with $6\leq Z \leq 20$
and $28\leq Z \leq 50$.

\begin{table*}[h] 
\caption{Ground-state properties of even-even isotopes with $Z=6-20$
obtained by the HF+BCS and HF with SkM$^*$; 
quadrupole deformation parameters $(\beta_2,\gamma)$,
pairing gaps for neutrons and protons $(\Delta_n,\Delta_p)$,
chemical potentials for neutrons and protons $(\lambda_n,\lambda_p)$.
In the case of normal phase ($\Delta=0$),
we define the chemical potential as the
single-particle energy of the highest occupied orbital,
$\lambda_n=\epsilon_N^0$ and $\lambda_p=\epsilon_Z^0$.
The pairing gaps and chemical potentials are given in units of MeV.
}
\label{tab: gs_properties}
\begin{center}
\begin{tabular}[t]{c|ccllrr|rcrr} \hline\hline
  & $\beta_2$ & $\gamma$ & $\Delta_n$ & $\Delta_p$ & $-\lambda_n$ & $-\lambda_p$ & $\ \beta_2^{\rm HF}$ & $\gamma^{\rm HF}$ & $-\lambda_n^{\rm HF}$ & $-\lambda_p^{\rm HF}$ 
\\ \hline
$^{ 8}$C  & 0.00 & $-$           & 0.00 & 0.00 & 31.22 & 2.96 & 0.00 & $-$         & 31.19 & 2.95\\
$^{10}$C & 0.23 & 0$^\circ$ & 0.00 & 0.00 & 17.00 & 7.86 & 0.23 & 0$^\circ$ & 16.99 & 7.84\\
$^{12}$C & 0.00 & $-$           & 0.00 & 0.00 & 16.82 & 14.07 & 0.00 & $-$         & 16.80 & 14.06\\
$^{14}$C & 0.00 & $-$           & 0.00 & 0.00 & 8.94 & 18.24 & 0.00 & $-$         & 8.94 & 18.23\\
$^{16}$C & 0.14 & 0$^\circ$ & 1.00 & 0.00 & 4.56 & 21.22 & 0.27 & 0$^\circ$ & 5.25 & 20.52\\
$^{18}$C & 0.27 & 0$^\circ$ & 0.57 & 0.00 & 3.90 & 23.71 & 0.30 & 0$^\circ$ & 4.69 & 23.53\\
$^{20}$C & 0.23 & 60$^\circ$ & 0.00 & 0.00 & 4.84 & 27.53 & 0.23 & 60$^\circ$ &4.83 & 27.50\\
$^{22}$C & 0.00 & $-$         & 0.00 & 0.00 & 3.41 & 30.25 & 0.00 & $-$          & 3.42 & 30.21\\
\hline
 $^{14}$O & 0.00 & $-$         & 0.00 & 0.00 & 20.81 & 5.69 & 0.00 &  $-$         & 20.80 &  5.68\\
 $^{16}$O & 0.00 & $-$         & 0.00 & 0.00 & 13.54 & 10.26 & 0.00 &  $-$         & 13.53 & 10.25\\
 $^{18}$O & 0.00 & $-$         & 1.06 & 0.00 & 7.94 & 13.91 & 0.15 & 0$^\circ$ & 7.75 & 13.58\\     
 $^{20}$O & 0.00 & $-$         & 1.10 & 0.00 & 7.45 & 17.34 & 0.18 & 0$^\circ$ & 7.80 & 16.81\\     
 $^{22}$O & 0.00 & $-$         & 0.00 & 0.00 & 8.06 & 20.66 & 0.00 &  $-$         & 8.04 & 20.65\\
 $^{24}$O & 0.00 & $-$         & 0.00 & 0.00 & 5.17 & 22.63 & 0.00 &  $-$         & 5.17 & 22.61\\
 $^{26}$O & 0.00 & $-$         & 0.80 & 0.00 & 1.13 & 24.87 & 0.07 & 60$^\circ$ & 1.26 & 24.79\\     
\hline
 $^{16}$Ne   & 0.25 & 0$^\circ$ & 0.00 & 0.54 & 23.06 & 0.91 & 0.29 & 0$^\circ$ & 22.81 & 1.48\\ 
 $^{18}$Ne   & 0.00 & $-$        & 0.00 & 1.07 & 17.16 & 4.01 & 0.16 & 0$^\circ$ & 16.82 & 3.86\\
 $^{20}$Ne   & 0.37 & 0$^\circ$  & 0.00  & 0.00 & 13.07 & 9.19 & 0.37 & 0$^\circ$ & 13.07 & 9.18 \\
 $^{22}$Ne   & 0.37 & 0$^\circ$  & 0.00 & 0.00 & 11.03 & 12.38 & 0.37 & 0$^\circ$ & 11.03 & 12.37 \\
 $^{24}$Ne   & 0.17 & 60$^\circ$ & 0.00 & 0.74 & 10.57 & 13.04 & 0.20 & 60$^\circ$ & 10.62 & 13.51\\
 $^{26}$Ne   & 0.00 & $-$        & 0.00 & 1.00 &  7.17 & 14.92 & 0.14 & 0$^\circ$ & 6.95 & 14.92\\
 $^{28}$Ne   & 0.00 & $-$        & 0.79 & 1.01 &  3.22 & 17.05 & 0.16 & 0$^\circ$ & 3.77 & 17.41\\
 $^{30}$Ne   & 0.00 & $-$        & 0.00 & 1.01 &  3.79 & 19.09 & 0.35 & 0$^\circ$ & 4.14 & 21.35\\
 $^{32}$Ne   & 0.36 & 0$^\circ$  & 0.95 & 0.00 &  2.16 & 23.61 & 0.41 & 0$^\circ$ & 2.92 & 24.28\\
\hline
 $^{18}$Mg & 0.31 & 0$^\circ$ & 0.00 & 0.00 & 25.59 & 0.20 & 0.31 & 0$^\circ$ & 25.56 & 0.19\\
 $^{20}$Mg & 0.00 & $-$            & 0.00 & 1.13 & 20.53 & 2.83 & 0.18 & 0$^\circ$ & 19.99 & 3.18\\
 $^{22}$Mg & 0.38 & 0$^\circ$ & 0.00 & 0.00 & 16.31 & 6.42 & 0.38 & 0$^\circ$ & 16.30 & 6.42\\
 $^{24}$Mg & 0.39 & 0$^\circ$  & 0.00 & 0.00 & 14.12 & 9.51 & 0.39 & 0$^\circ$ & 14.12 & 9.50\\
 $^{26}$Mg & 0.20 & 54$^\circ$ & 0.00 & 0.86 & 13.08 & 11.23 & 0.24 & 8$^\circ$ & 11.37 & 11.67\\
 $^{28}$Mg & 0.00 & $-$        & 0.00 & 1.03 &  9.21 & 13.30 & 0.19 & 0$^\circ$ & 8.99 & 13.80\\
 $^{30}$Mg & 0.00 & $-$        & 1.31 & 1.03 &  5.48 & 15.49 & 0.18 & 0$^\circ$ & 5.94 & 15.99\\
 $^{32}$Mg & 0.00 & $-$        & 0.00 & 1.03 &  5.83 & 17.55 & 0.35 & 0$^\circ$ & 6.08 & 18.08\\
 $^{34}$Mg & 0.37 & 0$^\circ$  & 1.45 & 0.00 &  4.12 & 20.18 & 0.41 & 0$^\circ$ & 5.10 & 20.31\\
 $^{36}$Mg & 0.33 & 0$^\circ$  & 1.43 & 0.00 &  3.21 & 21.95 & 0.31 & 0$^\circ$ & 4.03 & 21.83\\
 $^{38}$Mg & 0.30 & 0$^\circ$  & 1.47 & 0.00 &  2.38 & 23.69 & 0.30 & 8$^\circ$ & 3.16 & 23.87\\
 $^{40}$Mg & 0.29 & 0$^\circ$  & 0.91 & 0.00 &  1.31 & 25.28 & 0.32 & 0$^\circ$ & 2.48 & 25.33\\
\hline
  $^{24}$Si  &  0.18  &  60$^\circ$ &  0.71 &  0.00 &  16.94 &   5.26 & 0.20 &  60$^\circ$ &  17.41 &   5.30 \\
  $^{26}$Si  &  0.20  &  53$^\circ$ &  0.85 &  0.00 &  15.88 &   7.68 & 0.25 &  11$^\circ$ &  16.31 &   6.00 \\
  $^{28}$Si  &  0.23  &  60$^\circ$ &  0.00 &  0.00 &  15.72 &  10.36 & 0.23 &  60$^\circ$ &  15.69 &  10.33 \\
  $^{30}$Si  &  0.00  & $-$               &  0.00 &  0.00 &  11.25 &  12.74 & 0.00 &  $-$               &  11.24 &  12.73 \\
  $^{32}$Si  &  0.00  & $-$               &  1.31 &  0.00 &   7.63 &  14.92 & 0.13 &  60$^\circ$ &   8.08 &  14.55 \\
  $^{34}$Si  &  0.00  & $-$               &  1.65 &  0.00 &   6.57 &  16.95 & 0.00 &   $-$               &   7.87 &  16.98 \\
  $^{36}$Si  &  0.00  & $-$               &  2.16 &  0.00 &   5.61 &  18.82 & 0.12 &  60$^\circ$ &   5.70 &  18.50 \\
  $^{38}$Si  &  0.12  &  60$^\circ$ &  2.24 &  0.00 &   5.01 &  20.17 & 0.22 &   0$^\circ$ &   5.83 &  18.68 \\
  $^{40}$Si  &  0.17  &  60$^\circ$ &  2.03 &  0.00 &   4.35 &  21.82 & 0.22 &  22$^\circ$ &   5.07 &  20.65 \\
  $^{42}$Si  &  0.19  &  60$^\circ$ &  1.55 &  0.00 &   3.39 &  23.40 & 0.23 &  60$^\circ$ &   5.20 &  23.59 \\
  $^{44}$Si  &  0.00  & $-$               &  1.86 &  0.00 &   2.61 &  25.04 & 0.16 &  60$^\circ$ &   3.03 &  24.76 \\
  $^{46}$Si  &  0.00  & $-$               &  1.26 &  0.00 &   1.65 &  26.15 & 0.00 &   $-$               &   3.21 &  26.26 \\
\hline
\end{tabular}
\begin{tabular}[t]{c|ccllrr|ccrr} \hline\hline
  & $\beta_2$ & $\gamma$ & $\Delta_n$ & $\Delta_p$ & $-\lambda_n$ & $-\lambda_p$ & $\ \beta^{\rm HF}$ & $\gamma^{\rm HF}$ & $-\lambda_n^{\rm HF}$ & $-\lambda_p^{\rm HF}$ 
\\ \hline
 $^{26}$S & 0.00 & $-$        & 1.01 & 0.00 & 18.63 & 1.63 & 0.14 & 0$^\circ$ & 18.69 & 1.37\\
 $^{28}$S & 0.00 & $-$        & 1.03 & 0.00 & 17.85 & 3.49 & 0.20 & 0$^\circ$ & 18.35 & 3.25\\
 $^{30}$S & 0.00 & $-$        & 0.00 & 0.00 & 18.06 & 5.40 & 0.00 & $-$        & 18.04 & 5.39\\
 $^{32}$S & 0.00 & $-$        & 0.00 & 0.00 & 13.17 & 7.30 & 0.00 & $-$        & 13.15 & 7.29\\
 $^{34}$S & 0.00 & $-$        & 1.28 & 0.00 & 9.61 & 9.26 & 0.09 & 60$^\circ$ & 9.86 & 9.27\\
 $^{36}$S & 0.00 & $-$        & 1.52 & 0.00 & 8.36 & 11.08 & 0.00 & $-$        & 9.72 & 11.09\\
 $^{38}$S & 0.00 & $-$        & 2.19 & 0.00 & 7.34 & 12.77 & 0.15 & 0$^\circ$ & 7.50 & 13.14\\
 $^{40}$S & 0.16 & 0$^\circ$ & 2.07 & 0.00 & 6.73 & 14.94 & 0.23 & 0$^\circ$ & 7.77 & 15.24\\
 $^{42}$S & 0.15 & 0$^\circ$ & 2.06 & 0.00 & 5.82 & 16.66 & 0.22 & 0$^\circ$ & 6.65 & 17.13\\
 $^{44}$S & 0.11 & 0$^\circ$ & 1.97 & 0.00 & 4.88 & 18.00 & 0.10 & 0$^\circ$ & 5.89 & 18.13\\
 $^{46}$S & 0.04 & 1$^\circ$ & 1.72 & 0.00 & 3.97 & 18.97 & 0.15 & 0$^\circ$ & 4.57 & 19.74\\
 $^{48}$S & 0.00 & $-$        & 0.95 & 0.00 & 2.77 & 20.24 & 0.00 & $-$        & 4.50 & 20.31\\
 $^{50}$S & 0.00 & $-$        & 0.00 & 1.72 & 2.21 & 20.98 & 0.00 & $-$        & 2.19 & 21.09\\
\hline
 $^{32}$Ar & 0.00 & $-$        & 0.00 & 0.80 & 20.08 & 1.31 & 0.14 & 60$^\circ$ & 19.71 & 1.89\\
 $^{34}$Ar & 0.00 & $-$        & 0.00 & 0.80 & 15.11 & 3.25 & 0.08 & 60$^\circ$ & 15.10 & 3.59\\
 $^{36}$Ar & 0.00 & $-$        & 1.23 & 1.27 & 12.04 & 5.77 & 0.13 & 60$^\circ$ & 12.63 & 6.36\\
 $^{38}$Ar & 0.00 & $-$        & 1.11 & 1.24 & 10.48 & 8.03 & 0.04 & 60$^\circ$ & 11.60 & 8.10\\
 $^{40}$Ar & 0.00 & $-$        & 2.03 & 1.22 & 9.27 & 9.93 & 0.12 & 60$^\circ$ & 9.30 & 10.51\\
 $^{42}$Ar & 0.00 & $-$        & 2.16 & 1.20 & 8.43 & 11.79 & 0.18 & 0$^\circ$ & 9.17 & 11.73\\
 $^{44}$Ar & 0.00 & $-$        & 1.99 & 1.18 & 7.62 & 13.58 & 0.17 & 17$^\circ$ & 8.41 & 13.59\\
 $^{46}$Ar & 0.00 & $-$        & 1.77 & 1.16 & 6.49 & 15.20 & 0.14 & 60$^\circ$ & 8.17 & 16.04\\
 $^{48}$Ar & 0.00 & $-$        & 1.41 & 1.14 & 5.31 & 16.56 & 0.14 & 60$^\circ$ & 5.73 & 17.29\\
 $^{50}$Ar & 0.00 & $-$        & 0.85 & 2.00 & 3.95 & 17.62 & 0.08 & 60$^\circ$ & 5.01 & 18.13\\
 $^{52}$Ar & 0.01 & 59$^\circ$ & 1.22 & 0.00 & 2.37 & 19.02 & 0.06 & 60$^\circ$ & 3.08 & 19.03\\
 $^{54}$Ar & 0.12 & 60$^\circ$ & 0.00 & 0.00 & 2.22 & 20.96 & 0.12 & 60$^\circ$ & 2.22 & 20.93\\
 $^{56}$Ar & 0.00 & $-$        & 1.24 & 1.73 & 1.15 & 21.49 & 0.10 & 60$^\circ$ & 1.42 & 22.27\\
\hline
 $^{34}$Ca & 0.00 & $-$        & 0.00 & 0.00 & 21.99 & 1.13 & 0.00 & $-$        & 21.97 & 1.12\\
 $^{36}$Ca & 0.00 & $-$        & 0.00 & 0.00 & 16.92 & 2.93 & 0.00 & $-$        & 16.90 & 2.92\\
 $^{38}$Ca & 0.00 & $-$        & 1.19 & 1.20 & 14.29 & 3.64 & 0.04 & 60$^\circ$ & 14.36 & 4.76\\
 $^{40}$Ca & 0.00 & $-$        & 0.00 & 0.00 & 14.32 & 7.47 & 0.00 & $-$        & 14.30 & 7.46\\
 $^{42}$Ca & 0.00 & $-$        & 1.92 & 0.82 & 11.09 & 7.46 & 0.06 & 60$^\circ$ & 10.33 & 8.70\\
 $^{44}$Ca & 0.00 & $-$        & 2.05 & 0.00 & 10.22 & 11.13 & 0.10 & 0$^\circ$ & 10.41 & 10.25\\
 $^{46}$Ca & 0.00 & $-$        & 1.83 & 0.00 & 9.36 & 12.89 & 0.08 & 0$^\circ$ & 10.16 & 12.11\\
 $^{48}$Ca & 0.00 & $-$        & 0.00 & 0.00 & 10.41 & 14.69 & 0.00 & $-$        & 10.39 & 14.68\\
 $^{50}$Ca & 0.00 & $-$        & 1.46 & 1.90 & 6.53 & 13.77 & 0.08 & 60$^\circ$ & 6.59 & 15.40\\
 $^{52}$Ca & 0.00 & $-$        & 1.50 & 0.00 & 5.18 & 17.00 & 0.00 & $-$        & 6.67 & 16.95\\
 $^{54}$Ca & 0.00 & $-$        & 2.29 & 0.00 & 4.26 & 18.36 & 0.00 & $-$        & 4.11 & 17.99\\
 $^{56}$Ca & 0.00 & $-$        & 2.54 & 1.41 & 3.55 & 17.50 & 0.08 & 60$^\circ$ & 2.92 & 18.89\\
 $^{58}$Ca & 0.00 & $-$        & 2.54 & 0.00 & 3.01 & 21.12 & 0.08 & 0$^\circ$ & 3.06 & 20.59\\
 $^{60}$Ca & 0.00 & $-$        & 2.55 & 0.00 & 2.54 & 22.49 & 0.00 & $-$        & 3.40 & 22.36\\
 $^{62}$Ca & 0.00 & $-$        & 2.55 & 0.00 & 2.09 & 23.82 & 0.08 & 60$^\circ$ & 1.92 & 23.04\\
 $^{64}$Ca & 0.00 & $-$        & 2.55 & 0.75 & 1.60 & 22.44 & 0.08 & 0$^\circ$ & 2.23 & 24.15\\
\hline
\end{tabular}
\end{center}
\end{table*}

\begin{table*}[t]
\caption{Ground-state properties of Ni, Zn, Ge, Se and Kr isotopes 
obtained by the HF+BCS with SkM$^*$. 
See the caption of Table \ref{tab: gs_properties}.
The root mean square radii of neutrons and protons ($r_n,r_p$) are also shown.
}
\label{tab: gs_properties1}
\begin{center}
\begin{tabular}[t]{c|ccccrrrr} \hline\hline
  & $\beta_2$ & $\gamma$ & $r_n$ & $r_p$ & $\Delta_n$ & $\Delta_p$ & $-\lambda_n$ & $-\lambda_p$\\ 
\hline
  $^{56}$Ni & 0.00 & $-$   & 3.63 &  3.69 & 0.00 &  0.36 & 16.27 &  4.31\\
  $^{58}$Ni & 0.00 & $-$   & 3.70 &  3.70 & 1.50 &  0.55 & 11.32 &  5.58\\
  $^{60}$Ni & 0.00 & $-$   & 3.77 &  3.72 & 2.02 &  0.48 & 10.12 &  6.82\\
  $^{62}$Ni & 0.00 & $-$   & 3.84 &  3.75 & 2.23 &  0.00 &  9.16 & 10.70\\
  $^{64}$Ni & 0.00 & $-$   & 3.90 &  3.77 & 2.34 &  0.00 &  8.45 & 11.91\\
  $^{66}$Ni & 0.00 & $-$   & 3.96 &  3.79 & 2.41 &  0.00 &  7.86 & 13.11\\
  $^{68}$Ni & 0.00 & $-$   & 4.01 &  3.82 & 2.27 &  0.00 &  7.33 & 14.27\\
  $^{70}$Ni & 0.00 & $-$   & 4.06 &  3.84 & 2.22 &  0.00 &  6.87 & 15.38\\
  $^{72}$Ni & 0.00 & $-$   & 4.10 &  3.86 & 2.13 &  0.00 &  6.38 & 16.50\\
  $^{74}$Ni & 0.00 & $-$   & 4.14 &  3.87 & 1.98 &  0.00 &  5.89 & 17.58\\
  $^{76}$Ni & 0.00 & $-$   & 4.18 &  3.89 & 1.70 &  0.00 &  5.31 & 18.64\\
  $^{78}$Ni & 0.00 & $-$   & 4.21 &  3.90 & 0.86 &  0.00 &  4.32 & 19.69\\
  $^{80}$Ni & 0.00 & $-$   & 4.29 &  3.92 & 1.32 &  0.00 &  3.19 & 20.30\\
  $^{82}$Ni & 0.00 & $-$   & 4.36 &  3.93 & 1.36 &  0.00 &  2.65 & 20.92\\
  $^{84}$Ni & 0.00 & $-$   & 4.43 &  3.95 & 1.23 &  0.00 &  2.10 & 21.52\\
\hline
  $^{60}$Zn & 0.00 & $-$   & 3.72 &  3.78 & 1.43 &  0.96 & 12.59 &  3.03\\
  $^{62}$Zn & 0.01 & 51$^\circ$ & 3.79 &  3.80 & 1.89 &  1.21 & 11.29 &  4.28\\
  $^{64}$Zn & 0.00 & $-$   & 3.85 &  3.82 & 2.15 &  1.41 & 10.29 &  5.51\\
  $^{66}$Zn & 0.00 & $-$   & 3.91 &  3.85 & 2.28 &  1.51 &  9.57 &  6.67\\
  $^{68}$Zn & 0.00 & $-$   & 3.97 &  3.87 & 2.35 &  1.54 &  8.96 &  7.79\\
  $^{70}$Zn & 0.00 & $-$   & 4.02 &  3.89 & 2.20 &  1.58 &  8.40 &  8.89\\
  $^{72}$Zn & 0.00 & $-$   & 4.07 &  3.90 & 2.14 &  1.61 &  7.90 &  9.96\\
  $^{74}$Zn & 0.01 &  0$^\circ$ & 4.11 &  3.92 & 2.07 &  1.60 &  7.40 & 10.98\\
  $^{76}$Zn & 0.01 &  0$^\circ$ & 4.15 &  3.94 & 1.92 &  1.63 &  6.90 & 12.01\\
  $^{78}$Zn & 0.00 & $-$   & 4.19 &  3.95 & 1.62 &  1.73 &  6.33 & 13.08\\
  $^{80}$Zn & 0.00 & $-$   & 4.22 &  3.97 & 0.00 &  1.77 &  7.75 & 14.12\\
  $^{82}$Zn & 0.00 & $-$   & 4.29 &  3.98 & 1.08 &  1.68 &  3.96 & 14.85\\
  $^{84}$Zn & 0.01 &  0$^\circ$ & 4.35 &  3.99 & 1.11 &  1.61 &  3.48 & 15.60\\
  $^{86}$Zn & 0.00 & $-$   & 4.41 &  4.01 & 0.90 &  1.54 &  2.95 & 16.35\\
  $^{88}$Zn & 0.00 & $-$   & 4.48 &  4.02 & 0.58 &  1.47 &  2.03 & 17.02\\
\hline
  $^{64}$Ge & 0.00 & $-$   & 3.80 &  3.87 & 1.78 &  1.04 & 12.43 &  2.34\\
  $^{66}$Ge & 0.00 & $-$   & 3.87 &  3.89 & 2.09 &  1.13 & 11.40 &  3.53\\
  $^{68}$Ge & 0.04 & 59$^\circ$ & 3.93 &  3.92 & 2.26 &  1.74 & 10.72 &  4.95\\
  $^{70}$Ge & 0.00 & $-$   & 3.99 &  3.94 & 2.32 &  2.07 & 10.10 &  6.26\\
  $^{72}$Ge & 0.00 & $-$   & 4.04 &  3.95 & 2.13 &  2.10 &  9.51 &  7.40\\
  $^{74}$Ge & 0.00 & $-$   & 4.08 &  3.97 & 2.06 &  2.11 &  8.98 &  8.48\\
  $^{76}$Ge & 0.12 &  0$^\circ$ & 4.14 &  3.99 & 1.98 &  1.95 &  8.46 &  9.53\\
  $^{78}$Ge & 0.11 &  0$^\circ$ & 4.17 &  4.01 & 1.73 &  1.98 &  7.91 & 10.57\\
  $^{80}$Ge & 0.00 & $-$   & 4.20 &  4.01 & 1.54 &  2.16 &  7.38 & 11.60\\
  $^{82}$Ge & 0.00 & $-$   & 4.23 &  4.02 & 0.00 &  2.17 &  8.74 & 12.62\\
  $^{84}$Ge & 0.05 &  7$^\circ$ & 4.29 &  4.04 & 0.96 &  2.06 &  4.78 & 13.38\\
  $^{86}$Ge & 0.07 & 10$^\circ$ & 4.36 &  4.06 & 1.18 &  1.95 &  4.26 & 14.15\\
  $^{88}$Ge & 0.01 & 38$^\circ$ & 4.41 &  4.07 & 0.98 &  1.97 &  3.69 & 14.89\\
  $^{90}$Ge & 0.00 & $-$   & 4.48 &  4.08 & 1.03 &  1.92 &  2.83 & 15.57\\
  $^{92}$Ge & 0.17 & 28$^\circ$ & 4.55 &  4.14 & 1.72 &  1.57 &  2.79 & 16.58\\
  $^{94}$Ge & 0.22 & 26$^\circ$ & 4.61 &  4.18 & 1.75 &  1.36 &  2.59 & 17.43\\
  $^{96}$Ge & 0.25 & 22$^\circ$ & 4.67 &  4.21 & 1.71 &  1.29 &  2.34 & 18.24\\
  $^{98}$Ge & 0.26 & 18$^\circ$ & 4.73 &  4.24 & 1.82 &  1.29 &  2.06 & 18.98\\
\hline
\end{tabular}
\begin{tabular}[t]{c|ccccrrrr} 
\hline\hline
  & $\beta_2$ & $\gamma$ & $r_n$ & $r_p$ & $\Delta_n$ & $\Delta_p$ & $-\lambda_n$ & $-\lambda_p$ \\
\hline
  $^{68}$Se & 0.09 &  0$^\circ$ & 3.90 &  3.97 & 2.09 &  1.13 & 12.67 &  2.03\\
  $^{70}$Se & 0.18 & 60$^\circ$ & 3.98 &  4.01 & 2.09 &  1.57 & 11.94 &  3.60\\
  $^{72}$Se & 0.14 & 60$^\circ$ & 4.02 &  4.02 & 2.13 &  1.98 & 11.20 &  4.83\\
  $^{74}$Se & 0.00 & $-$   & 4.05 &  4.02 & 2.08 &  2.19 & 10.63 &  5.98\\
  $^{76}$Se & 0.01 & 46$^\circ$ & 4.10 &  4.03 & 1.99 &  2.22 & 10.06 &  7.08\\
  $^{78}$Se & 0.10 &  0$^\circ$ & 4.15 &  4.05 & 1.93 &  2.15 &  9.53 &  8.09\\
  $^{80}$Se & 0.10 &  0$^\circ$ & 4.18 &  4.06 & 1.68 &  2.16 &  8.96 &  9.12\\
  $^{82}$Se & 0.00 & $-$   & 4.21 &  4.07 & 1.47 &  2.27 &  8.43 & 10.20\\
  $^{84}$Se & 0.00 & $-$   & 4.24 &  4.08 & 0.00 &  2.29 &  9.73 & 11.21\\
  $^{86}$Se & 0.04 &  0$^\circ$ & 4.30 &  4.09 & 1.13 &  2.21 &  5.65 & 11.97\\
  $^{88}$Se & 0.07 &  0$^\circ$ & 4.36 &  4.11 & 1.32 &  2.10 &  5.08 & 12.72\\
  $^{90}$Se & 0.06 & 60$^\circ$ & 4.41 &  4.13 & 1.22 &  2.07 &  4.41 & 13.49\\
  $^{92}$Se & 0.10 & 30$^\circ$ & 4.48 &  4.15 & 1.65 &  1.94 &  3.93 & 14.26\\
  $^{94}$Se & 0.20 & 59$^\circ$ & 4.56 &  4.21 & 1.63 &  1.54 &  3.76 & 15.16\\
  $^{96}$Se & 0.33 &  4$^\circ$ & 4.63 &  4.28 & 1.39 &  1.56 &  3.59 & 16.26\\
  $^{98}$Se & 0.32 &  9$^\circ$ & 4.68 &  4.30 & 1.51 &  1.52 &  3.23 & 16.96\\
 $^{100}$Se & 0.30 &  0$^\circ$ & 4.72 &  4.30 & 1.64 &  1.56 &  2.89 & 17.63\\
 $^{102}$Se & 0.27 &  0$^\circ$ & 4.76 &  4.31 & 1.76 &  1.51 &  2.56 & 18.26\\
 $^{104}$Se & 0.25 &  0$^\circ$ & 4.80 &  4.33 & 1.74 &  1.46 &  2.23 & 18.91\\
\hline
  $^{72}$Kr & 0.26 & 60$^\circ$ & 4.03 &  4.11 & 1.78 &  0.97 & 13.09 &  2.15\\
  $^{74}$Kr & 0.14 & 60$^\circ$ & 4.04 &  4.08 & 2.09 &  2.09 & 12.32 &  3.55\\
  $^{76}$Kr & 0.00 & $-$   & 4.07 &  4.08 & 2.03 &  2.28 & 11.74 &  4.72\\
  $^{78}$Kr & 0.00 & $-$   & 4.11 &  4.09 & 1.92 &  2.29 & 11.13 &  5.80\\
  $^{80}$Kr & 0.00 & $-$   & 4.15 &  4.10 & 1.86 &  2.31 & 10.56 &  6.84\\
  $^{82}$Kr & 0.00 & $-$   & 4.19 &  4.11 & 1.72 &  2.32 & 10.03 &  7.86\\
  $^{84}$Kr & 0.00 & $-$   & 4.22 &  4.12 & 1.40 &  2.33 &  9.46 &  8.86\\
  $^{86}$Kr & 0.00 & $-$   & 4.25 &  4.13 & 0.00 &  2.35 & 10.70 &  9.85\\
  $^{88}$Kr & 0.01 & 25$^\circ$ & 4.31 &  4.14 & 1.18 &  2.26 &  6.44 & 10.62\\
  $^{90}$Kr & 0.03 &  0$^\circ$ & 4.37 &  4.16 & 1.41 &  2.18 &  5.88 & 11.39\\
  $^{92}$Kr & 0.00 & $-$   & 4.42 &  4.18 & 1.64 &  2.15 &  5.32 & 12.18\\
  $^{94}$Kr & 0.06 & 44$^\circ$ & 4.47 &  4.20 & 1.63 &  2.04 &  4.68 & 12.92\\
  $^{96}$Kr & 0.18 & 60$^\circ$ & 4.55 &  4.25 & 1.74 &  1.69 &  4.57 & 13.64\\
  $^{98}$Kr & 0.36 &  0$^\circ$ & 4.64 &  4.35 & 1.37 &  1.41 &  4.43 & 14.93\\
 $^{100}$Kr & 0.23 & 60$^\circ$ & 4.66 &  4.31 & 1.76 &  1.47 &  3.95 & 15.18\\
 $^{102}$Kr & 0.35 &  0$^\circ$ & 4.73 &  4.38 & 1.33 &  1.42 &  3.56 & 16.42\\
 $^{104}$Kr & 0.32 &  1$^\circ$ & 4.77 &  4.39 & 1.54 &  1.43 &  3.23 & 17.02\\
 $^{106}$Kr & 0.31 &  0$^\circ$ & 4.80 &  4.40 & 1.51 &  1.43 &  2.87 & 17.63\\
 $^{108}$Kr & 0.28 &  0$^\circ$ & 4.84 &  4.41 & 1.48 &  1.42 &  2.55 & 18.18\\
 $^{110}$Kr & 0.17 & 32$^\circ$ & 4.84 &  4.38 & 1.86 &  1.57 &  2.50 & 18.88\\
 $^{112}$Kr & 0.06 & 13$^\circ$ & 4.84 &  4.38 & 1.78 &  1.78 &  2.54 & 19.90\\
 $^{114}$Kr & 0.00 & $-$   & 4.86 &  4.39 & 1.52 &  1.84 &  2.40 & 20.74\\
 $^{116}$Kr & 0.00 & $-$   & 4.89 &  4.41 & 1.21 &  1.80 &  2.05 & 21.45\\
 $^{118}$Kr & 0.00 & $-$   & 4.92 &  4.43 & 0.00 &  1.77 &  3.72 & 22.14\\
\hline
\end{tabular}
\end{center}
\end{table*}

\begin{table*}[t]
\caption{
Same as Table \ref{tab: gs_properties1}, but for Sr, Zr, Mo, and Ru isotopes. 
}
\begin{center}
\begin{tabular}[t]{c|ccccrrrr} \hline\hline
  & $\beta_2$ & $\gamma$ & $r_n$ & $r_p$ & $\Delta_n$ & $\Delta_p$ & $-\lambda_n$ & $-\lambda_p$\\ 
\hline
  $^{76}$Sr & 0.16 & 60$^\circ$ & 4.06 &  4.14 & 2.05 &  1.59 & 13.41 &  2.17\\
  $^{78}$Sr & 0.00 & $-$   & 4.09 &  4.13 & 1.98 &  2.26 & 12.82 &  3.51\\
  $^{80}$Sr & 0.00 & $-$   & 4.13 &  4.14 & 1.85 &  2.25 & 12.17 &  4.55\\
  $^{82}$Sr & 0.00 & $-$   & 4.17 &  4.15 & 1.81 &  2.25 & 11.59 &  5.57\\
  $^{84}$Sr & 0.00 & $-$   & 4.20 &  4.16 & 1.67 &  2.25 & 11.04 &  6.57\\
  $^{86}$Sr & 0.00 & $-$   & 4.23 &  4.17 & 1.35 &  2.24 & 10.46 &  7.54\\
  $^{88}$Sr & 0.00 & $-$   & 4.26 &  4.18 & 0.00 &  2.24 & 11.65 &  8.50\\
  $^{90}$Sr & 0.00 & $-$   & 4.32 &  4.19 & 1.29 &  2.16 &  7.30 &  9.28\\
  $^{92}$Sr & 0.00 & $-$   & 4.37 &  4.21 & 1.49 &  2.10 &  6.68 & 10.06\\
  $^{94}$Sr & 0.00 & $-$   & 4.42 &  4.22 & 1.66 &  2.05 &  6.10 & 10.84\\
  $^{96}$Sr & 0.00 & $-$   & 4.47 &  4.24 & 1.67 &  2.00 &  5.48 & 11.61\\
  $^{98}$Sr & 0.37 &  0$^\circ$ & 4.60 &  4.38 & 1.34 &  1.21 &  5.85 & 12.44\\
 $^{100}$Sr & 0.38 &  0$^\circ$ & 4.65 &  4.40 & 1.27 &  1.11 &  5.26 & 13.28\\
 $^{102}$Sr & 0.38 &  0$^\circ$ & 4.70 &  4.42 & 1.39 &  1.06 &  4.77 & 14.07\\
 $^{104}$Sr & 0.37 &  0$^\circ$ & 4.74 &  4.44 & 1.41 &  1.05 &  4.29 & 14.80\\
 $^{106}$Sr & 0.36 &  0$^\circ$ & 4.78 &  4.45 & 1.41 &  1.06 &  3.88 & 15.50\\
 $^{108}$Sr & 0.35 &  0$^\circ$ & 4.81 &  4.47 & 1.25 &  1.06 &  3.48 & 16.19\\
 $^{110}$Sr & 0.34 &  0$^\circ$ & 4.85 &  4.48 & 1.31 &  1.11 &  3.11 & 16.81\\
 $^{112}$Sr & 0.14 & 60$^\circ$ & 4.81 & 4.41 & 1.57 &  1.51 &  3.31 & 17.65\\
 $^{114}$Sr & 0.00 & $-$   & 4.83 &  4.41 & 1.61 &  1.66 &  3.44 & 18.56\\
 $^{116}$Sr & 0.00 & $-$   & 4.86 &  4.43 & 1.56 &  1.63 &  3.15 & 19.24\\
 $^{118}$Sr & 0.00 & $-$   & 4.88 &  4.44 & 1.16 &  1.59 &  2.91 & 19.95\\
\hline
  $^{80}$Zr & 0.00 & $-$   & 4.11 &  4.19 & 1.93 &  1.92 & 13.87 &  2.32\\
  $^{82}$Zr & 0.00 & $-$   & 4.15 &  4.19 & 1.80 &  1.91 & 13.18 &  3.34\\
  $^{84}$Zr & 0.00 & $-$   & 4.18 &  4.20 & 1.76 &  1.91 & 12.58 &  4.33\\
  $^{86}$Zr & 0.00 & $-$   & 4.22 &  4.21 & 1.63 &  1.90 & 12.02 &  5.30\\
  $^{88}$Zr & 0.00 & $-$   & 4.25 &  4.21 & 1.31 &  1.90 & 11.44 &  6.25\\
  $^{90}$Zr & 0.00 & $-$   & 4.28 &  4.22 & 0.00 &  1.89 & 10.00 &  7.18\\
  $^{92}$Zr & 0.00 & $-$   & 4.33 &  4.24 & 1.31 &  1.82 &  8.10 &  7.95\\
  $^{94}$Zr & 0.00 & $-$   & 4.38 &  4.25 & 1.54 &  1.76 &  7.47 &  8.71\\
  $^{96}$Zr & 0.00 & $-$   & 4.43 &  4.27 & 1.68 &  1.71 &  6.88 &  9.48\\
  $^{98}$Zr & 0.00 & $-$   & 4.48 &  4.28 & 1.65 &  1.65 &  6.25 & 10.24\\
 $^{100}$Zr & 0.36 &  0$^\circ$ & 4.60 &  4.41 & 1.35 &  1.35 &  6.61 & 10.76\\
 $^{102}$Zr & 0.37 &  0$^\circ$ & 4.65 &  4.44 & 1.32 &  1.26 &  6.06 & 11.48\\
 $^{104}$Zr & 0.38 &  0$^\circ$ & 4.70 &  4.46 & 1.33 &  1.16 &  5.53 & 12.20\\
 $^{106}$Zr & 0.38 &  0$^\circ$ & 4.74 &  4.48 & 1.27 &  1.07 &  5.01 & 12.92\\
 $^{108}$Zr & 0.37 &  0$^\circ$ & 4.77 &  4.50 & 1.34 &  1.10 &  4.60 & 13.66\\
 $^{110}$Zr & 0.36 &  0$^\circ$ & 4.81 &  4.51 & 1.28 &  1.11 &  4.20 & 14.38\\
 $^{112}$Zr & 0.35 &  0$^\circ$ & 4.85 &  4.52 & 1.21 &  1.13 &  3.77 & 15.08\\
 $^{114}$Zr & 0.16 & 52$^\circ$ & 4.81 &  4.46 & 1.56 &  1.36 &  3.98 & 16.24\\
 $^{116}$Zr & 0.00 & $-$   & 4.83 &  4.45 & 1.69 &  1.36 &  4.14 & 16.99\\
 $^{118}$Zr & 0.00 & $-$   & 4.85 &  4.46 & 1.50 &  1.35 &  3.90 & 17.69\\
 $^{120}$Zr & 0.00 & $-$   & 4.88 &  4.48 & 1.18 &  1.34 &  3.60 & 18.37\\
 $^{122}$Zr & 0.00 & $-$   & 4.90 &  4.49 & 0.00 &  1.35 &  2.93 & 19.07\\
 $^{124}$Zr & 0.00 & $-$   & 4.97 &  4.50 & 1.12 &  1.28 &  1.58 & 19.43\\
 $^{126}$Zr & 0.00 & $-$   & 5.02 &  4.51 & 1.26 &  1.23 &  1.23 & 19.83\\
 $^{128}$Zr & 0.00 & $-$   & 5.08 &  4.52 & 1.31 &  1.17 &  0.91 & 20.23\\
 $^{130}$Zr & 0.00 & $-$   & 5.12 &  4.53 & 0.84 &  1.11 &  0.77 & 20.60\\
\hline
\end{tabular}
\begin{tabular}[t]{c|ccccrrrr} 
\hline\hline
  & $\beta_2$ & $\gamma$ & $r_n$ & $r_p$ & $\Delta_n$ & $\Delta_p$ & $-\lambda_n$ & $-\lambda_p$ \\
\hline
  $^{84}$Mo & 0.00 & $-$   & 4.16 &  4.24 & 1.76 &  1.69 & 14.17 &  2.17\\
  $^{86}$Mo & 0.00 & $-$   & 4.19 &  4.25 & 1.72 &  1.70 & 13.55 &  3.14\\
  $^{88}$Mo & 0.00 & $-$   & 4.23 &  4.25 & 1.59 &  1.72 & 12.98 &  4.09\\
  $^{90}$Mo & 0.00 & $-$   & 4.26 &  4.26 & 1.28 &  1.72 & 12.40 &  5.02\\
  $^{92}$Mo & 0.00 & $-$   & 4.29 &  4.26 & 0.00 &  1.74 & 13.48 &  5.95\\
  $^{94}$Mo & 0.00 & $-$   & 4.34 &  4.28 & 1.32 &  1.69 &  8.88 &  6.69\\
  $^{96}$Mo & 0.00 & $-$   & 4.39 &  4.29 & 1.58 &  1.64 &  8.24 &  7.44\\
  $^{98}$Mo & 0.00 & $-$   & 4.43 &  4.31 & 1.71 &  1.59 &  7.64 &  8.20\\
 $^{100}$Mo & 0.00 & $-$   & 4.48 &  4.33 & 1.67 &  1.55 &  7.02 &  8.95\\
 $^{102}$Mo & 0.29 & 12$^\circ$ & 4.58 &  4.42 & 1.55 &  1.43 &  7.20 &  9.67\\
 $^{104}$Mo & 0.32 & 12$^\circ$ & 4.63 &  4.45 & 1.42 &  1.33 &  6.81 & 10.33\\
 $^{106}$Mo & 0.33 & 14$^\circ$ & 4.68 &  4.48 & 1.36 &  1.27 &  6.33 & 11.07\\
 $^{108}$Mo & 0.32 & 17$^\circ$ & 4.72 &  4.49 & 1.35 &  1.22 &  5.87 & 11.87\\
 $^{110}$Mo & 0.30 & 19$^\circ$ & 4.75 &  4.50 & 1.38 &  1.22 &  5.44 & 12.66\\
 $^{112}$Mo & 0.28 & 23$^\circ$ & 4.78 &  4.51 & 1.38 &  1.26 &  5.07 & 13.47\\
 $^{114}$Mo & 0.24 & 27$^\circ$ & 4.80 &  4.51 & 1.39 &  1.30 &  4.79 & 14.26\\
 $^{116}$Mo & 0.20 & 33$^\circ$ & 4.82 &  4.51 & 1.32 &  1.29 &  4.56 & 14.96\\
 $^{118}$Mo & 0.00 & $-$   & 4.82 &  4.48 & 1.64 &  1.36 &  4.86 & 15.67\\
 $^{120}$Mo & 0.00 & $-$   & 4.85 &  4.50 & 1.46 &  1.35 &  4.61 & 16.37\\
 $^{122}$Mo & 0.00 & $-$   & 4.87 &  4.51 & 1.14 &  1.36 &  4.32 & 17.06\\
 $^{124}$Mo & 0.00 & $-$   & 4.90 &  4.53 & 0.00 &  1.36 &  5.75 & 17.76\\
 $^{126}$Mo & 0.00 & $-$   & 4.95 &  4.54 & 0.70 &  1.31 &  1.78 & 18.19\\
 $^{128}$Mo & 0.00 & $-$   & 5.00 &  4.55 & 0.98 &  1.27 &  1.63 & 18.56\\
 $^{130}$Mo & 0.00 & $-$   & 5.05 &  4.56 & 0.97 &  1.22 &  1.37 & 18.95\\
 $^{132}$Mo & 0.00 & $-$   & 5.10 &  4.57 & 0.89 &  1.19 &  1.08 & 19.32\\
\hline
  $^{88}$Ru & 0.00 & $-$   & 4.21 &  4.29 & 1.69 &  1.64 & 14.50 &  2.05\\
  $^{90}$Ru & 0.00 & $-$   & 4.24 &  4.29 & 1.56 &  1.65 & 13.93 &  2.99\\
  $^{92}$Ru & 0.00 & $-$   & 4.27 &  4.30 & 1.25 &  1.66 & 13.33 &  3.90\\
  $^{94}$Ru & 0.00 & $-$   & 4.30 &  4.30 & 0.00 &  1.67 & 14.38 &  4.81\\
  $^{96}$Ru & 0.00 & $-$   & 4.35 &  4.32 & 1.33 &  1.61 &  9.64 &  5.56\\
  $^{98}$Ru & 0.00 & $-$   & 4.40 &  4.34 & 1.62 &  1.60 &  9.00 &  6.30\\
 $^{100}$Ru & 0.00 & $-$   & 4.44 &  4.35 & 1.74 &  1.56 &  8.40 &  7.06\\
 $^{102}$Ru & 0.00 & $-$   & 4.48 &  4.37 & 1.72 &  1.53 &  7.80 &  7.81\\
 $^{104}$Ru & 0.13 &  0$^\circ$ & 4.54 &  4.40 & 1.85 &  1.52 &  7.58 &  8.73\\
 $^{106}$Ru & 0.27 & 18$^\circ$ & 4.62 &  4.47 & 1.60 &  1.19 &  7.49 &  9.07\\
 $^{108}$Ru & 0.28 & 19$^\circ$ & 4.66 &  4.49 & 1.53 &  1.13 &  7.08 &  9.76\\
 $^{110}$Ru & 0.28 & 21$^\circ$ & 4.70 &  4.51 & 1.49 &  1.08 &  6.66 & 10.51\\
 $^{112}$Ru & 0.27 & 23$^\circ$ & 4.74 &  4.52 & 1.45 &  1.04 &  6.24 & 11.28\\
 $^{114}$Ru & 0.25 & 26$^\circ$ & 4.77 &  4.53 & 1.41 &  1.02 &  5.85 & 12.05\\
 $^{116}$Ru & 0.23 & 27$^\circ$ & 4.79 &  4.54 & 1.38 &  1.04 &  5.51 & 12.81\\
 $^{118}$Ru & 0.04 &  1$^\circ$ & 4.79 &  4.51 & 1.76 &  1.37 &  5.75 & 13.85\\
 $^{120}$Ru & 0.00 & $-$   & 4.82 &  4.52 & 1.63 &  1.36 &  5.56 & 14.54\\
 $^{122}$Ru & 0.00 & $-$   & 4.84 &  4.53 & 1.43 &  1.36 &  5.31 & 15.24\\
 $^{124}$Ru & 0.00 & $-$   & 4.87 &  4.55 & 1.11 &  1.35 &  5.02 & 15.93\\
 $^{126}$Ru & 0.00 & $-$   & 4.89 &  4.56 & 0.00 &  1.36 &  6.44 & 16.62\\
 $^{128}$Ru & 0.00 & $-$   & 4.94 &  4.57 & 0.75 &  1.31 &  2.29 & 17.04\\
 $^{130}$Ru & 0.00 & $-$   & 4.99 &  4.59 & 0.83 &  1.29 &  2.04 & 17.46\\
 $^{132}$Ru & 0.00 & $-$   & 5.04 &  4.60 & 0.89 &  1.25 &  1.80 & 17.85\\
 $^{134}$Ru & 0.00 & $-$   & 5.09 &  4.61 & 0.80 &  1.23 &  1.48 & 18.24\\
\hline
\end{tabular}
\end{center}
\end{table*}

\begin{table*}[t]
\caption{
Same as Table \ref{tab: gs_properties1}, but for Pd and Cd isotopes.
}
\begin{center}
\begin{tabular}[t]{c|ccccrrrr} \hline\hline
  & $\beta_2$ & $\gamma$ & $r_n$ & $r_p$ & $\Delta_n$ & $\Delta_p$ & $-\lambda_n$ & $-\lambda_p$\\ 
\hline
  $^{92}$Pd & 0.00 & $-$   & 4.25 &  4.33 & 1.54 &  1.49 & 14.85 &  1.94\\
  $^{94}$Pd & 0.00 & $-$   & 4.28 &  4.34 & 1.23 &  1.49 & 14.25 &  2.84\\
  $^{96}$Pd & 0.00 & $-$   & 4.31 &  4.34 & 0.00 &  1.50 & 15.27 &  3.73\\
  $^{98}$Pd & 0.00 & $-$   & 4.36 &  4.36 & 1.34 &  1.46 & 10.39 &  4.48\\
 $^{100}$Pd & 0.00 & $-$   & 4.40 &  4.37 & 1.65 &  1.42 &  9.75 &  5.24\\
 $^{102}$Pd & 0.00 & $-$   & 4.45 &  4.39 & 1.76 &  1.43 &  9.15 &  5.99\\
 $^{104}$Pd & 0.08 &  0$^\circ$ & 4.50 &  4.41 & 1.82 &  1.33 &  8.68 &  6.75\\
 $^{106}$Pd & 0.14 &  0$^\circ$ & 4.55 &  4.44 & 1.85 &  1.19 &  8.37 &  7.51\\
 $^{108}$Pd & 0.17 &  0$^\circ$ & 4.60 &  4.47 & 1.87 &  1.13 &  8.05 &  8.23\\
 $^{110}$Pd & 0.19 &  0$^\circ$ & 4.64 &  4.49 & 1.85 &  1.10 &  7.72 &  8.93\\
 $^{112}$Pd & 0.20 &  0$^\circ$ & 4.68 &  4.51 & 1.79 &  1.08 &  7.38 &  9.64\\
 $^{114}$Pd & 0.19 &  0$^\circ$ & 4.71 &  4.52 & 1.71 &  1.07 &  7.04 & 10.39\\
 $^{116}$Pd & 0.18 &  0$^\circ$ & 4.74 &  4.53 & 1.65 &  1.06 &  6.71 & 11.17\\
 $^{118}$Pd & 0.15 &  0$^\circ$ & 4.77 &  4.54 & 1.62 &  1.06 &  6.46 & 11.95\\
 $^{120}$Pd & 0.10 &  0$^\circ$ & 4.79 &  4.55 & 1.66 &  1.11 &  6.34 & 12.73\\
 $^{122}$Pd & 0.00 & $-$   & 4.81 &  4.55 & 1.62 &  1.26 &  6.26 & 13.47\\
 $^{124}$Pd & 0.00 & $-$   & 4.84 &  4.57 & 1.43 &  1.25 &  6.00 & 14.17\\
 $^{126}$Pd & 0.00 & $-$   & 4.87 &  4.58 & 1.10 &  1.24 &  5.69 & 14.85\\
 $^{128}$Pd & 0.00 & $-$   & 4.89 &  4.59 & 0.00 &  1.24 &  7.12 & 15.53\\
 $^{130}$Pd & 0.00 & $-$   & 4.94 &  4.61 & 0.71 &  1.21 &  2.72 & 15.96\\
 $^{132}$Pd & 0.00 & $-$   & 4.98 &  4.62 & 0.80 &  1.19 &  2.47 & 16.40\\
 $^{134}$Pd & 0.00 & $-$   & 5.03 &  4.63 & 0.75 &  1.16 &  2.23 & 16.84\\
 $^{136}$Pd & 0.00 & $-$   & 5.07 &  4.64 & 0.60 &  1.15 &  1.86 & 17.25\\
\hline
\end{tabular}
\begin{tabular}[t]{c|ccccrrrr} 
\hline\hline
  & $\beta_2$ & $\gamma$ & $r_n$ & $r_p$ & $\Delta_n$ & $\Delta_p$ & $-\lambda_n$ & $-\lambda_p$ \\
\hline
  $^{96}$Cd & 0.00 & $-$   & 4.29 &  4.38 & 1.21 &  1.17 & 15.16 &  1.79\\
  $^{98}$Cd & 0.00 & $-$   & 4.32 &  4.38 & 0.00 &  1.18 & 16.14 &  2.66\\
 $^{100}$Cd & 0.00 & $-$   & 4.37 &  4.40 & 1.35 &  1.14 & 11.13 &  3.42\\
 $^{102}$Cd & 0.00 & $-$   & 4.41 &  4.41 & 1.67 &  1.11 & 10.49 &  4.19\\
 $^{104}$Cd & 0.00 & $-$   & 4.45 &  4.43 & 1.78 &  1.13 &  9.90 &  4.92\\
 $^{106}$Cd & 0.00 & $-$   & 4.50 &  4.44 & 1.80 &  1.11 &  9.36 &  5.69\\
 $^{108}$Cd & 0.09 &  0$^\circ$ & 4.55 &  4.47 & 1.91 &  0.91 &  9.07 &  6.31\\
 $^{110}$Cd & 0.13 &  0$^\circ$ & 4.59 &  4.49 & 1.91 &  0.76 &  8.75 &  6.89\\
 $^{112}$Cd & 0.15 &  0$^\circ$ & 4.63 &  4.51 & 1.91 &  0.64 &  8.43 &  7.52\\
 $^{114}$Cd & 0.16 &  0$^\circ$ & 4.67 &  4.53 & 1.87 &  0.55 &  8.11 &  8.17\\
 $^{116}$Cd & 0.16 &  0$^\circ$ & 4.70 &  4.55 & 1.82 &  0.49 &  7.79 &  8.88\\
 $^{118}$Cd & 0.15 &  0$^\circ$ & 4.74 &  4.56 & 1.75 &  0.50 &  7.49 &  9.66\\
 $^{120}$Cd & 0.14 &  0$^\circ$ & 4.77 &  4.57 & 1.64 &  0.00 &  7.19 & 11.33\\
 $^{122}$Cd & 0.00 & $-$   & 4.78 &  4.57 & 1.75 &  1.00 &  7.19 & 11.72\\
 $^{124}$Cd & 0.00 & $-$   & 4.81 &  4.59 & 1.62 &  0.99 &  6.94 & 12.41\\
 $^{126}$Cd & 0.00 & $-$   & 4.84 &  4.60 & 1.42 &  0.98 &  6.67 & 13.10\\
 $^{128}$Cd & 0.00 & $-$   & 4.87 &  4.61 & 1.10 &  0.98 &  6.36 & 13.78\\
 $^{130}$Cd & 0.00 & $-$   & 4.89 &  4.63 & 0.00 &  0.97 &  7.80 & 14.45\\
 $^{132}$Cd & 0.00 & $-$   & 4.93 &  4.64 & 0.64 &  0.95 &  3.13 & 14.91\\
 $^{134}$Cd & 0.00 & $-$   & 4.98 &  4.65 & 0.72 &  0.94 &  2.90 & 15.36\\
 $^{136}$Cd & 0.00 & $-$   & 5.02 &  4.66 & 0.76 &  0.92 &  2.66 & 15.80\\
 $^{138}$Cd & 0.00 & $-$   & 5.06 &  4.68 & 0.59 &  0.91 &  2.24 & 16.23\\
\hline
\end{tabular}
\end{center}
\end{table*}

\begin{table*}[t]
Ground-state properties of Sn isotopes calculated with SkM$^*$ and SkI3.
See the captions of Tables \ref{tab: gs_properties} and \ref{tab: gs_properties1}. 
\label{tab: gs_properties_Sn}
\begin{center}
\begin{tabular}[t]{c|ccccrrrr} \hline\hline
SkM$^*$ & $\beta_2$ & $\gamma$ & $r_n$ & $r_p$ & $\Delta_n$ & $\Delta_p$ & $-\lambda_n$ & $-\lambda_p$ \\
 \hline
 $^{100}$Sn & 0.00 & $-$             & 4.33 & 4.42 & 0.00 & 0.00 & 17.00 & 3.15\\
 $^{102}$Sn & 0.00 & $-$             & 4.38 & 4.43 & 1.35 & 0.00 & 11.86 & 3.86\\
 $^{104}$Sn & 0.00 & $-$             & 4.42 & 4.45 & 1.70 & 0.00 & 11.24 & 4.59\\
 $^{106}$Sn & 0.00 & $-$             & 4.46 & 4.46 & 1.80 & 0.00 & 10.65 & 5.33\\
 $^{108}$Sn & 0.00 & $-$             & 4.50 & 4.48 & 1.84 & 0.00 & 10.13 & 6.07\\
 $^{110}$Sn & 0.00 & $-$             & 4.54 & 4.49 & 1.97 & 0.00 & 9.75 & 6.84\\
 $^{112}$Sn & 0.00 & $-$             & 4.58 & 4.51 & 2.05 & 0.00 & 9.43 & 7.60\\
 $^{114}$Sn & 0.00 & $-$             & 4.62 & 4.53 & 2.10 & 0.00 & 9.13 & 8.35\\
 $^{116}$Sn & 0.00 & $-$             & 4.66 & 4.55 & 2.06 & 0.00 & 8.86 & 9.09\\
 $^{118}$Sn & 0.00 & $-$             & 4.69 & 4.56 & 2.00 & 0.00 & 8.60 & 9.82\\
 $^{120}$Sn & 0.00 & $-$             & 4.72 & 4.58 & 1.94 & 0.00 & 8.36 & 10.54\\
 $^{122}$Sn & 0.00 & $-$             & 4.75 & 4.59 & 1.85 & 0.00 & 8.12 & 11.25\\
 $^{124}$Sn & 0.00 & $-$             & 4.78 & 4.61 & 1.75 & 0.00 & 7.87 & 11.95\\
 $^{126}$Sn & 0.00 & $-$             & 4.81 & 4.62 & 1.62 & 0.00 & 7.62 & 12.63\\
 $^{128}$Sn & 0.00 & $-$             & 4.84 & 4.63 & 1.42 & 0.00 & 7.34 & 13.31\\
 $^{130}$Sn & 0.00 & $-$             & 4.87 & 4.64 & 1.10 & 0.00 & 7.01 & 13.98\\
 $^{132}$Sn & 0.00 & $-$             & 4.89 & 4.66 & 0.00 & 0.00 & 8.48 & 14.64\\
 $^{134}$Sn & 0.00 & $-$             & 4.93 & 4.67 & 0.68 & 0.00 & 3.62 & 15.07\\
 $^{136}$Sn & 0.00 & $-$             & 4.97 & 4.68 & 0.78 & 0.00 & 3.36 & 15.51\\
 $^{138}$Sn & 0.00 & $-$             & 5.02 & 4.69 & 1.04 & 0.00 & 3.11 & 15.90\\
 $^{140}$Sn & 0.00 & $-$             & 5.07 & 4.71 & 1.74 & 0.00 & 3.06 & 16.43\\
\hline
\end{tabular}
\begin{tabular}[t]{c|ccccrrrr} \hline \hline
SkI3 & $\beta_2$ & $\gamma$ & $r_n$ & $r_p$ & $\Delta_n$ & $\Delta_p$ & $-\lambda_n$ & $-\lambda_p$ \\
\hline 
 $^{100}$Sn & 0.00 & $-$             & 4.33 & 4.40 & 0.00 & 0.00 & 17.18 & 3.50\\
 $^{102}$Sn & 0.00 & $-$             & 4.38 & 4.42 & 1.54 & 0.00 & 11.00 & 4.33\\
 $^{104}$Sn & 0.00 & $-$             & 4.43 & 4.43 & 1.65 & 0.00 & 10.40 & 5.16\\
 $^{106}$Sn & 0.00 & $-$             & 4.48 & 4.45 & 2.20 & 0.00 & 10.34 & 6.00\\
 $^{108}$Sn & 0.00 & $-$             & 4.52 & 4.47 & 1.83 & 0.00 & 9.93 & 6.81\\
 $^{110}$Sn & 0.00 & $-$             & 4.56 & 4.49 & 2.37 & 0.00 & 9.76 & 7.61\\
 $^{112}$Sn & 0.00 & $-$             & 4.60 & 4.50 & 2.43 & 0.00 & 9.41 & 8.38\\
 $^{114}$Sn & 0.00 & $-$             & 4.64 & 4.52 & 2.31 & 0.00 & 9.09 & 9.13\\
 $^{116}$Sn & 0.00 & $-$             & 4.67 & 4.53 & 2.02 & 0.00 & 8.69 & 9.85\\
 $^{118}$Sn & 0.00 & $-$             & 4.71 & 4.55 & 2.04 & 0.00 & 8.28 & 10.55\\
 $^{120}$Sn & 0.00 & $-$             & 4.75 & 4.56 & 0.00 & 0.00 & 8.50 & 10.94\\
 $^{122}$Sn & 0.00 & $-$             & 4.78 & 4.57 & 1.70 & 0.00 & 7.61 & 11.94\\
 $^{124}$Sn & 0.00 & $-$             & 4.81 & 4.59 & 1.68 & 0.00 & 7.34 & 12.67\\
 $^{126}$Sn & 0.00 & $-$             & 4.84 & 4.60 & 1.61 & 0.00 & 7.10 & 13.40\\
 $^{128}$Sn & 0.00 & $-$             & 4.87 & 4.61 & 1.46 & 0.00 & 6.85 & 14.13\\
 $^{130}$Sn & 0.00 & $-$             & 4.90 & 4.62 & 1.19 & 0.00 & 6.54 & 14.86\\
 $^{132}$Sn & 0.00 & $-$             & 4.93 & 4.64 & 0.00 & 0.00 & 8.00 & 15.56\\
 $^{134}$Sn & 0.00 & $-$             & 4.99 & 4.65 & 0.78 & 0.00 & 2.00 & 15.93\\
 $^{136}$Sn & 0.00 & $-$             & 5.04 & 4.67 & 1.30 & 0.00 & 2.10 & 16.39\\
 $^{138}$Sn & 0.00 & $-$             & 5.10 & 4.69 & 1.61 & 0.00 & 2.09 & 16.63\\
 $^{140}$Sn & 0.00 & $-$             & 5.14 & 4.71 & 1.66 & 0.00 & 1.94 & 17.22\\
\hline
\end{tabular}
\end{center}
\end{table*}

\bibliography{nuclear_physics,myself,current}

\end{document}